\documentclass{emulateapj}
\usepackage{epstopdf}
\usepackage{placeins}
\usepackage{graphicx}
\usepackage{amsmath}
\shorttitle{Spiral Density Wakes} \shortauthors{Zhu et al.}

\newcommand\be{\begin{equation}}
\newcommand\en{\end{equation}}

\newcommand\etal{{\rm et al}.\ }

\begin{document}

\title{THE STRUCTURE OF SPIRAL SHOCKS EXCITED BY PLANETARY-MASS COMPANIONS}

\author{Zhaohuan Zhu\altaffilmark{1,4}, Ruobing Dong\altaffilmark{2,3,4},   
James M. Stone\altaffilmark{1}, and Roman R. Rafikov\altaffilmark{1}}

\altaffiltext{1}{Department of Astrophysical Sciences, 4 Ivy Lane, Peyton Hall,
Princeton University, Princeton, NJ 08544, USA}
\altaffiltext{2}{Lawrence Berkeley National Lab, Berkeley, CA 94720, rdong2013@berkeley.edu}
\altaffiltext{3}{Department of Astronomy, University of California at Berkeley, Berkeley, CA 94720}
\altaffiltext{4}{Hubble Fellow.}

\email{zhzhu@astro.princeton.edu }

\begin{abstract}
Direct imaging observations have revealed spiral structures in protoplanetary disks. Previous studies have suggested that  planet-induced spiral arms cannot explain some of these spiral patterns, due to the large pitch angle and high contrast of the spiral arms in observations. We have carried out three dimensional (3-D) hydrodynamical simulations to study spiral wakes/shocks excited by young planets. We find that, in contrast with linear theory, the pitch angle of spiral arms does depend on the planet mass, which can be explained by  the non-linear density wave theory. A secondary {(or even a tertiary)} spiral arm, especially for inner arms, is also excited by a massive planet. With a more massive planet in the disk, the excited spiral arms have larger pitch angle and the separation between the primary and secondary arms in the azimuthal direction is also larger. We also find that although the arms in the outer disk do not exhibit much vertical motion, the inner arms have significant vertical motion, which boosts the density perturbation at the disk atmosphere. Combining hydrodynamical models with Monte-Carlo radiative transfer calculations, we find that the inner spiral arms are considerably more prominent in synthetic near-IR images using full 3-D hydrodynamical models than images based on 2-D models assuming vertical hydrostatic equilibrium, indicating the need to model observations with full 3-D hydrodynamics. Overall, companion-induced spiral arms not only pinpoint the companion's position but also provide three independent ways (pitch angle, separation between two arms, and contrast of arms) to constrain the companion's mass.  
 \end{abstract}


\section{Introduction}
Recent high-resolution direct imaging observations have revealed spiral structure in {three
protoplanetary disks around Herbig Ae/Be stars: SAO 206462 (Muto \etal 2012; Garufi et al. 2013), MWC 758
(Grady \etal 2013; Benisty \etal 2015), and HD 100546 (Currie et al. 2014).}
The polarized intensity has been measured in these observations to gain higher 
contrast between the disk and the central star. While the thermal emission from the central star is  
unpolarized, the scattered light from the disk
is polarized. In these near-infrared (near-IR) polarized intensity images,  two spiral arms
with roughly 180$^o$ rotational symmetry are present {in both SAO 206462 and MWC 758}, 
similar to the grand design in a spiral galaxy (e.g. the Whirlpool Galaxy M51). 
The spiral arms also exhibit a high contrast against the background disk.
The polarized intensity of the spiral arm is several times higher than that of 
the region outside the spiral arm.
It should also be noted that, since
the dust scattering opacity is quite large, these observations only probe structure 
high up at the disk atmosphere (e.g. several disk scale heights) where the last dust scattering surface is.

 In addition to spiral patterns,  these {three} disks also have gaps or holes {(SAO 206462 with a submillimeter cavity of 46 AU, MWC 758 with a cavity of 73 AU from Andrews et al. 2011,
 and HD 100546 with a cavity of 10 AU revealed by SED fitting from Bouwman et 
al. 2003)}, 
 which indicates that they are members of the protoplanetary disk class called transitional disks (Espaillat \etal 2014).
One scenario to explain both spiral patterns and gaps/holes is that these disks harbor low-mass companions (e.g. young planets) which
can open gaps and excite spiral waves at the same time (e.g., Baruteau \etal 2014). 

However, there are two difficulties in explaining 
the observed spiral patterns using planet-induced spiral wakes.
First, the large pitch angle of {all} the observed 
spiral arms suggests that the disk has a relative high temperature 
(e.g. $\sim$250 K at 70 AU for MWC 758, Benisty \etal 2015). 
In linear theory, spiral waves are basically sound waves in disks, and  the pitch angle of the spiral arms is directly related to the sound speed in the disk. Using the linear theory, the 
best fit models for both SAO 206462 (Muto \etal 2012) and MWC 758 (Grady \etal 2013; Benisty \etal 2015)
suggest that the disk aspect ratio ($H/R$ with $H\equiv c_{s}/\Omega$) at $R\sim 100$ AU is around 0.2  which is too large for any realistic disk structure. For example, even if the stellar irradiation is perpendicular to the disk surface \footnote{In reality,
the stellar irradiation to the disk is not that efficient since the light from the star impinges very obliquely on the disk.}, 
the maximum disk temperature due to the stellar irradiation is $\sigma T(R)^{4}=L_{*}/4\pi R^2$ so that $H/R\propto T^{1/2} \propto L_{*}^{1/8}$. 
Assuming a  2 $M_{\odot}$ central star with 10 $L_{\odot}$ luminosity, 
the maximum temperature  is $\sim$70 K at 100 AU and $H/R$ is only $\sim$0.1. 
Since $H/R\propto L_{*}^{1/8}$, it is very difficult to make $H/R\sim$0.2. 

Second, the observed spiral arms exhibit much higher brightness contrasts than suggested by the synthetic
observations based on two dimensional (2-D) planet-disk simulations. Juh{\'a}sz \etal (2015) have calculated the polarized scattered  light images by combining 2-D hydrodynamical simulations with 3-D Monte-Carlo radiative transfer (MCRT) simulations. Vertical hydrostatic equilibrium has been assumed to extend the 2-D simulation to the third dimension (the vertical direction).
 They find that a relative change of about 3.5 on the spiral 
 arms in the surface density  is required 
for the spirals to be detectable.  This value is a factor of eight higher than what is seen in their hydrodynamical simulations. 

In this paper, we first point out that the pitch angle formula derived from 
the linear theory, which has been used in almost all previous
spiral arm modeling efforts, does not apply to the high planet mass cases. 
Spiral wakes that are excited by high mass planets (e.g. 1 $M_{J}$)
become spiral shocks which propagate at speeds faster
than the local sound speed {(Goodman \& Rafikov 2001, Rafikov 2002)}. Thus, the  pitch angle difficulty above 
can be alleviated by considering the non-linear 
extension of the spiral shock theory.  We also show that spiral arms  (especially the inner arms)
have complicated non-hydrostatic 3-D structure. Such structure can lead to strong density 
perturbation at the disk surface resulting in a corrugated shape of its atmosphere. 
Since near-IR observations only probe the shape of disk surface, this effect alleviates the second difficulty mentioned above.  In Dong \etal (2015), we have combined MCRT
simulations with hydrodynamical simulations from this paper and demonstrated that planet-induced 
inner spiral arms can explain recent near-IR direct imaging observations of SAO 206462 and MWC 758.
{We note that, since the planets that we have proposed are outside the spiral arms, we cannot explain the gaps discovered at small radii in these transitional disks. 
Other mechanisms, e.g. another
planet or photoevaporation,  are needed to explain these gaps.}

Before we introduce our numerical method
in \S 3, we provide the theoretical background in \S 2. 
The shape of the spiral wakes will be studied in \S 4, and their 3-D structure will be presented
in \S 5. After a short discussion in \S 6, we summarize our results in \S 7. 

\section{Theoretical Background}
 As a result of  planet-disk interaction, a spiral arm forms due to the 
constructive interference of density wakes with different  azimuthal wavenumbers $m$ excited by the planet at Lindblad resonances. 
In the linear density wave theory, the $m$-th
Fourier component of the planet potential  excites 
the density wave having $m$ spiral arms
\begin{equation}
\delta(R,\phi,t)=\delta_{0}(R)e^{i[\int k_{R}(R)dR + m(\phi-\Omega_{p}t)]}
\end{equation}
where $\delta$ is 
any perturbed quantity associated with the wave, $\delta_{0}(R)$ is its complex amplitude,
 $k_{R}(R)$ is the radial wave vector, and $\Omega_{p}$ is the planet orbital frequency. Thus, the wave has the same phase along the curve satisfying $dR/d\phi=-m/k_{R}(R)$.
The pitch angle ($\beta$) of the equal phase curve satisfies tan$\beta=|dR/(R d\phi)|$, so $\beta$=tan$^{-1} |m/[k_{R}(R)R]|$.
Using the dispersion relationship for density waves in the large $m$ limit and far from the launching point, $m^2(\Omega(R)-\Omega_{p})^2\approx c_{s}^2 k_{R}(R)^{2}$,  
we have $\beta$=tan$^{-1}[c_{s}/(R|\Omega(R)-\Omega_{p}|)]$. Because $\beta$ is independent of $m$, different $m$ modes can constructively
interfere to form the one armed spiral wake (Ogilvie \& Lubow 2002). If the equal phase curve is integrated from the planet's position ($R_{p}$, $\phi_p$), the shape of the wake far from $R_{p}$ is given by Rafikov (2002) and Muto \etal (2012) as
\begin{eqnarray}
\phi(R)&=&\phi_{p}-\frac{{\rm sgn}(R-R_{p})}{h_{p}} \nonumber \\
 &&\times\left[\left(\frac{R}{R_{p}}\right)^{1+\eta}\left\{\frac{1}{1+\eta}-\frac{1}{1-\alpha+\eta}\left(\frac{R}{R_{p}}\right)^{-\alpha}\right\}\right. \nonumber \\
 &&\,\,\,\,\,\,\,\,\,\left.-\left(\frac{1}{1+\eta}-\frac{1}{1-\alpha+\eta}\right)\right]\label{eq:linearwake}
\end{eqnarray}
{where $h_{p}=H/R$ is the disk aspect ratio at $R_{p}$, $\Omega(R)\propto R^{-\alpha}$, and the sound speed $c_{s}(R)\propto R^{-\eta}$}.

However, when the planet is massive enough, the above linear density wave theory breaks down.
Linear waves excited by planets will steepen
to shocks (Goodman \& Rafikov 2001, Rafikov 2002, Dong, Rafikov \& Stone  2012, Duffell 
\& MacFadyen 2012, Zhu \etal 2013) 
after they propagate over a distance
\begin{equation}
|x_{sh}|\approx 0.93 \left(\frac{\gamma+1}{12/5}\frac{M_{p}}{M_{th}}\right)^{-2/5} H\,.\label{eq:eqshock}
\end{equation} 
where $\gamma$ is the adiabatic index, and $M_{th}$ 
is the disk thermal mass
\begin{equation}
M_{th}\equiv\frac{c_{s}^{3}}{G\Omega_{p}}\approx 1
M_{J}\left(\frac{h_{p}}{0.1 }\right)^{3}\left(\frac{M_{*}}{M_{\odot}}\right)\,.\label{eq:thermal}
\end{equation} 
When $M_{p}>M_{th}$, the spiral waves will immediately become spiral shocks after they are excited around the planet.
Unlike the linear wake which follows Equation (\ref{eq:linearwake}), the spiral shock will expand 
away from the trajectory predicted by Equation (\ref{eq:linearwake}). Thus, if there is a massive planet in  the disk, using 
 Equation (\ref{eq:linearwake}) to fit the shape of the spiral shocks
will predict an incorrect disk aspect ratio and temperature.

\section{Numerical Simulations}
\subsection{Method}
To study density wakes/shocks excited by planets, we have carried out both 2-D and 3-D {hydrodynamical}  simulations using Athena++.
Athena++ is a newly developed grid based code using a higher-order Godunov scheme for MHD and 
the constrained transport (CT) to conserve the divergence-free property for magnetic fields (Stone \etal,  in preparation). 
{But in this paper, we do not include magnetic fields and only solve hydrodynamical equations using Athena++.}
Compared with its predecessor Athena 
(Gardiner \& Stone 2005, 2008; Stone \etal 2008),  Athena++ is highly optimized and uses flexible grid structures, allowing global
numerical simulations spanning a large radial range. Furthermore, the geometric source terms in curvilinear coordinates 
(e.g. in cylindrical and spherical-polar coordinates) are carefully implemented so that angular momentum
is conserved exactly (to machine precision), which makes the code ideal for global disk simulations. 

  Our simulations use the adiabatic equation of state (EoS) with
the adiabatic index $\gamma$=1.4.  A simple orbital cooling scheme has been applied
to mimic the radiative cooling process in disks. In 3-D simulations, we have adopted
\begin{equation}
\frac{dE}{dt}=-\frac{E-c_{v}\rho T_{irr}}{t_{cool}}\,,
\end{equation}
where $\rho$ and $E$ are the density and the internal energy per unit volume, while in 2-D simulations, we have adopted
\begin{equation}
\frac{dE}{dt}=-\frac{E-c_{v}\Sigma T_{irr}}{t_{cool}}\label{eq:dEdt}
\end{equation}
where $\Sigma$ and $E$ are the disk surface density and 
the internal energy per unit area. $c_{v}\equiv k/(\mu m_{u} (\gamma-1))$ is the heat capacity per unit mass, $k$ is the Boltzmann constant, $\mu$ is the mean molecular weight, and $m_{u}$ is the atomic mass unit. The cooling time
$t_{cool}$ can be written in the dimensionless form as
$T_{cool}=t_{cool}\Omega(R)$. We fix $T_{cool}$ to be
a constant in each simulation. With this scheme, the disk temperature is
relaxed to the background disk temperature ($T_{irr}$)
determined by stellar irradiation. In our simulations, $T_{irr}$ is set to be the initial disk temperature.
{To estimate $T_{cool}$ in a realistic disk, we use the grey atmosphere approximation (Hubeny 1990) for the radiative cooling,}
\begin{equation}
\frac{dE}{dt}=-\frac{16}{3}\sigma(T_{c}^4-T_{irr}^4)\frac{\tau}{1+\tau^2}\,,\label{eq:dEdt2}
\end{equation}
where $\sigma$ is the Stefan-Boltzmann constant, $\tau=(\Sigma/2) \kappa_{R}$ is the optical depth in the vertical direction, $\kappa_{R}$ is the Rosseland mean opacity, 
and $T_{c}$ is the midplane temperature. Assuming $E=c_{v}\Sigma T_{c}$ and using Equations
(\ref{eq:dEdt}) and (\ref{eq:dEdt2}), we can derive
\begin{equation}
t_{cool}=\frac{3\Sigma c_{v}}{16\sigma (T_{c}^{2}+T_{irr}^2)(T_{c}+T_{irr})}\frac{1+\tau^{2}}{\tau}\,.\label{eq:tcool}
\end{equation}
Approximating the polynomial of $T_{c}$ and $T_{irr}$ in the denominator of Equation \ref{eq:tcool}
with $[{\rm max}(T_{c}, T_{irr})]^3$ and assuming the
 central star is 1 $M_{\odot}$, we have 
\begin{eqnarray}
T_{cool}&=&0.002\left(\frac{\Sigma}{10 {\rm g cm}^{-2}}\right)\left(\frac{100 {\rm AU}}{R}\right)^{1.5}
\frac{(60 {\rm K})^3}{[{\rm max}(T_{c}, T_{irr})]^3}\nonumber\\
&&\times \frac{1+\tau^2}{\tau}\label{eq:tcoolvalue}
\end{eqnarray}

{Thus, $T_{cool}$ can vary dramatically at different radii in disks. Using the minimum mass solar nebular model, $T_{cool}$ is $\sim$10$^{5}$ at 1 AU and $10^{-2}$
at 100 AU.} We have carried out three sets of simulations
with $T_{cool}=10^{-5}$, 1, and 100. They are respectively labeled as ISO, T1, T2 at the end of their names in Table 1. Simulations with fast cooling ($T_{cool}=10^{-5}$) are equivalent to locally
isothermal simulations.
{Simulations with $T_{cool}=100$ are basically adiabatic simulations considering the timescale of the simulations is several tens of orbits, and simulations with $T_{c}=1$ should be between isothermal and adiabatic simulations. Somewhat surprisingly, 
we find that simulations with $T_{cool}=1$ are
qualitatively similar to those with $T_{cool}=100$. 
This similarity suggests that the spiral waves in disks with $T_{cool}=1$ behave adiabatically. 
We think that this is due to the short timescale for the flow to move across the spiral wake.  The spiral wake has a typical width smaller than the disk scale height, and
the background flow moves across the wake at nearly Keplerian speed. Thus, when the fluid travels in and out of the wave/shock, its response time is much 
smaller than the orbital time
and should behave adiabatically even in disks with $T_{cool}=1$.}
Considering this similarity, in most part of the paper, we only show results with $T_{cool}=10^{-5}$ and 1.

\begin{table}[ht]
\begin{center}
\caption{Models \label{tab1}}
\begin{tabular}{cccccccc}
\tableline
\tableline
2-D &&&&&& \\
\tableline
\tableline
Run    & $M_{p}$  & $T_{cool}$ & Domain & Resolution &  \\
 & $M_{J}$   &  & $R$    &  $R\times\phi$ \\
\tableline
CM1ISO  & 0.01  & 10$^{-5}$ & [0.2,10] & 1280$\times$2048  \\
CM1T1  & 0.01  & 1 & [0.2,10] & 1280$\times$2048  \\
CM1T100  & 0.01  & 100 & [0.2,10] & 1280$\times$2048  \\
CM2ISO  &  1  & 10$^{-5}$ & [0.2,10] & 1280$\times$2048  \\
CM2T1  &  1  & 1 & [0.2,10] & 1280$\times$2048  \\
CM2T100  & 1  & 100 & [0.2,10] & 1280$\times$2048  \\
CM3ISO  &  6  & 10$^{-5}$ & [0.2,10] & 1280$\times$2048  \\
CM3T1  &  6  & 1 & [0.2,10] & 1280$\times$2048  \\
CM3T100  & 6  & 100 & [0.2,10] & 1280$\times$2048  \\
\tableline
\tableline
3-D &&&&&& \\
\tableline
\tableline
Run    & $M_{p}$  & $T_{cool}$ & Domain & Resolution &  \\
 & $M_{J}$   &   & $r\times\theta$    &  $r\times\theta\times\phi$ \\
\tableline
STHIN & 0.000316 &  10$^{-5}$ & [0.5,2]$\times$[$\frac{\pi}{2}$-$0.2$,$\frac{\pi}{2}$+$0.2$] & 456$\times$128$\times$2048  \\
SM1ISO  & 0.01  & 10$^{-5}$ & [0.3,3]$\times$[$\frac{\pi}{2}$-$0.6$,$\frac{\pi}{2}$+$0.6$] & 256$\times$128$\times$688  \\
SM1T1  & 0.01  & 1 & [0.3,3]$\times$[$\frac{\pi}{2}$-$0.6$,$\frac{\pi}{2}$+$0.6$]  & 256$\times$128$\times$688  \\
SM1T100  & 0.01  & 100 & [0.3,3]$\times$[$\frac{\pi}{2}$-$0.6$,$\frac{\pi}{2}$+$0.6$]  & 256$\times$128$\times$688  \\
SM2ISO  &  1  & 10$^{-5}$ & [0.3,3]$\times$[$\frac{\pi}{2}$-$0.6$,$\frac{\pi}{2}$+$0.6$] & 256$\times$128$\times$688  \\
SM2T1  &  1  & 1 & [0.3,3]$\times$[$\frac{\pi}{2}$-$0.6$,$\frac{\pi}{2}$+$0.6$]  & 256$\times$128$\times$688  \\
SM2T100  & 1  & 100 & [0.3,3]$\times$[$\frac{\pi}{2}$-$0.6$,$\frac{\pi}{2}$+$0.6$]  & 256$\times$128$\times$688  \\
SM3ISO  &  6  & 10$^{-5}$ & [0.3,3]$\times$[$\frac{\pi}{2}$-$0.6$,$\frac{\pi}{2}$+$0.6$]  & 256$\times$128$\times$688  \\
SM3T1  &  6  & 1 & [0.3,3]$\times$[$\frac{\pi}{2}$-$0.6$,$\frac{\pi}{2}$+$0.6$]  & 256$\times$128$\times$688  \\
SM3T100  & 6  & 100 & [0.3,3]$\times$[$\frac{\pi}{2}$-$0.6$,$\frac{\pi}{2}$+$0.6$]  & 256$\times$128$\times$688  \\
\tableline
SM2ISOlong &  1  & 10$^{-5}$ & [0.3,3]$\times$[$\frac{\pi}{2}$-$0.6$,$\frac{\pi}{2}$+$0.6$] & 256$\times$128$\times$688  \\
SM2ISOcyl &  1  & 10$^{-5}$ & [0.3,3]$\times$[-0.8,0.8] \tablenotemark{a} & 296$\times$176$\times$690 \tablenotemark{b} \\
SM2ISOconstT &  1  & 10$^{-5}$ & [0.3,3]$\times$[$\frac{\pi}{2}$-$0.6$,$\frac{\pi}{2}$+$0.6$] & 256$\times$128$\times$688  \\
\tableline
\end{tabular}
\tablenotetext{1}{ This is the domain size in the $R\times Z$ direction since cylindrical coordinates
have been used.}
\tablenotetext{1}{ The resolution is for $R\times Z \times \phi$ with the cylindrical coordinate system.}
\end{center}
\end{table}

We have also varied the planet mass to be 0.01, 1, and 6 $M_{J}$ in our main set of simulations, which 
are labeled as M1, M2, and M3 in their names respectively. {Here, we have defined $M_{J}$ as 0.001 of the central star's mass.}
The thermal mass (Equation \ref{eq:thermal}) for the $h=0.1$ disk is $\sim M_{J}$. Thus, waves excited by a 0.01 $M_{J}$ planet
are in the linear regime and waves from a 6 $M_{J}$ planet are in the highly non-linear regime. 
To compare with Figure 2 in Tanaka \etal (2002), we have also carried out a thin disk simulation with $H_{p}/R_{p}=10^{-1.5}$ (STHIN in Table 1). The thermal mass for such a thin disk is only 
$M_{th}=$0.0316 $M_{J}$. Thus, in order to ensure that the waves are in the linear regime, we choose the planet mass of $0.01 M_{th}=3.16\times 10^{-4} M_{J}$ in this thin disk simulation.
To avoid the divergence of planet potential, a smoothing length of  0.1 $R_{p}$ has been applied for M2 and M3 cases ({0.1 $R_{p}$ is close to the disk scale height and the Hill radius of the planet}). 
For the thin disk case which has a very small mass planet, 
we choose a smoothing length of 6$\times$10$^{-3} R_{p}$, roughly the length of two grid cells.
For the low mass planet cases (M1),
a smoothing length of 0.02 $R_{p}$, which is also roughly the length of two grid cells in these simulations, has been adopted.
Planets
are fixed in circular orbits at $R=1$, and the indirect potential, which is due to the center
of the coordinate system is at the star instead of the center of the mass, has been included.
We have run the simulations for 10 planetary orbits. We choose this timescale because 
it is longer than the sound crossing time throughout the whole disk so that density waves/shocks
have established, while it is shorter than the gap opening timescale to avoid complicated
gap structures (e.g. vortices at the gap edges) and other longterm effects (e.g. radial buoyancy
waves, Richert et al. 2015). {To verify that  the revealed wave mechanics still hold in long terms we have run
one simulation for 120 orbits (SM2ISOlong in Table 1), which will be discussed in \S 6.1. }
A constant $\alpha$ 
viscosity with $\alpha=10^{-4}$ has been applied in our main sets of simulations. 

{At inner and outer boundaries, all quantities are fixed at the initial states. For a numerical code using Godunov scheme
which calculates the flux by decomposing wave characteristics at the left and right grid cells, 
such boundary can absorb wave characteristics coming from
the active zones and limit  waves traveling from the ghost zones to active zones.  Such boundary condition shows little wave reflection and 
is similar to the non-reflecting boundary condition (Godon 1996) used in  the FARGO code (Masset 2000). The detailed code comparison  
is given in Appendix C of Zhu \etal 2014.}

\subsection{2-D simulations}
Compared with the 3-D simulations in the next subsection, 2-D simulations allow us
to study density wakes in a bigger domain using a higher numerical resolution. The initial radial profile of the disk is
\begin{eqnarray}
\Sigma_{0}(R,\phi)=\Sigma_{0}(R_{0})\left(\frac{R}{R_{0}}\right)^{-1} \label{eq:eqinid}\\
T_{0}(R,\phi)=T_{0}(R_{0})\left(\frac{R}{R_{0}}\right)^{-1/2}\,.\label{eq:eqinit}
\end{eqnarray}
We choose $R_{0}=1$, $\Sigma_{0}(R_{0})=1$, and $T_{0}(R_{0})=0.01$ to make $(H/R)_{R=R_{0}}=0.1$.

Cylindrical coordinates have been adopted. To make every grid cell have equal length in the 
radial and azimuthal direction throughout the whole domain,  the grids are uniformly
spaced in log($R$) from $R=$0.2 to 10, and uniformly 
spaced from $0$ to $2\pi$ in the  $\phi$ direction. Our standard resolution 
is 1280 in the $R$ direction and $2048$ in the $\phi$ direction, which is equivalent to 32 grids per $H$ at $R=1$
in both directions. In Table 1, 2-D runs are denoted with a ``C''  in front of the model names, while 3-D runs are denoted with a ``S''  in front of the model names.

\subsection{3-D simulations}
To study the 3-D structure of  density wakes/shocks, we have run  3-D hydrodynamical simulations in spherical polar coordinates {except for one case in cylindrical coordinates}. 
The initial density profile of the disk at the disk midplane is
\begin{equation}
\rho_{0}(R,z=0)=\rho_{0}(R_{0},z=0)\left(\frac{R}{R_{0}}\right)^p\,,
\end{equation}
and the temperature is constant on cylinders
\begin{equation}
T_{0}(R,z)=T_{0}(R_{0})\left(\frac{R}{R_{0}}\right)^q\,.
\end{equation}
We want to emphasize that $R$ should not be confused with $r$.
In this paper, we use ($R$, $\phi$, $z$) to represent positions in cylindrical coordinates while using 
($r$, $\theta$, $\phi$) for spherical polar coordinates. 
$\phi$ represents the azimuthal direction (the direction of disk rotation) in both coordinate systems. Considering the disk structure
is more natural to be described in cylindrical coordinates, we have transformed 3-D simulation results from spherical polar coordinates  to cylindrical coordinates. Most results presented below
are  plotted in cylindrical coordinates with $R$ representing the distance to the axis of the disk, even though most simulations are carried out in spherical polar coordinates. 

Hydrostatic equilibrium in the $r-\theta$ plane requires that (e.g. Nelson \etal 2013)
\begin{equation}
\rho_{0}(R,z)=\rho_{0}(R_{0},z=0)\left(\frac{R}{R_{0}}\right)^p {\rm exp}\left[\frac{GM}{c_{s}^2}\left(\frac{1}{\sqrt{R^2+z^2}}-\frac{1}{R}\right)\right]\,,
\end{equation}
and
\begin{equation}
\Omega(R,z)=\Omega_{K}\left[(p+q)\left(\frac{H}{R}\right)^2+(1+q)-\frac{qR}{\sqrt{R^2+z^2}}\right]^{1/2}\,,
\end{equation}
where $c_{s}=\sqrt{p/\rho}$ is the isothermal sound speed at $R$, $\Omega_{K}=\sqrt{GM_{*}/R^3}$, and $H=c_{s}/\Omega_{K}$ as defined before.

We choose $p=-2.25$ and $q=-1/2$ in our main sets of simulations so that $\Sigma_{0}\propto R^{-1}$, similar to 2-D simulations above.
{$\rho_{0}(R_{0},z=0)$  is 1} and $H/R$ is 0.1 at $R=R_{0}$.
The grids are uniformly spaced in log($r$),  $\theta$, $\phi$ with 256$\times$128$\times$688 grid cells
in the domain of [log(0.3), log(3)]$\times$[$\pi/2$-0.6, $\pi/2$+0.6 ]$\times$[0, 2$\pi$] for the main sets of simulations.  
In  runs with $T_{cool}=1$ and 100, the cooling time decreases exponentially beyond $z=3 H$ with $T_{cool}(z)=T_{cool}{\rm exp}(-(z^2/H^2-3^2))$ to mimic fast cooling at the disk surface (D'Alessio et al. 1998). Numerically, this treatment also maintains better hydrostatic equilibrium at the disk surface.

The boundary condition in the $\theta$ direction is chosen that $v_{r}=v_{\theta}=0$ in the ghost zones.  
We set $v_{\phi}$ and $T$ in the ghost zones
having the same values as the last active zones. Density in the ghost zones is set to be
\begin{equation}
\rho(\theta_{g})=\rho(\theta_{a})\left|\frac{{\rm sin}(\theta_{g})}{{\rm sin}(\theta_{a})}\right|^{v_{\phi}^2/T}
\end{equation}
to maintain hydrostatic equilibrium in the $\theta$ direction, 
where $\theta_{g}$ and $\theta_{a}$ are the $\theta$ coordinates of the ghost and last active zones.
{We have also tried the boundary condition which sets the quantities in the ghost zones as the initial values,
and found that the results are not affected by the choice of boundary conditions. }

{To further study the numerical effect from the boundary, we have applied a wave damping zone (de Val-Borro et al. 2006)  operating at both $r$ and $\theta$
boundaries in the run SM2ISOlong. The wave damping zone in the $r$ direction is from $R_{in}$ to 1.25$R_{in}$ and from 0.84$R_{out}$ to $R_{out}$.
The damping zone in the $\theta$ direction is from $\theta_{in}$ to $\theta_{in}+0.1$ and from $\theta_{out}-0.1$ to $\theta_{out}$.
And $R_{in}$, $R_{out}$, $\theta_{in}$, and $\theta_{out}$ are the boundary of the simulation domain.  In these damping zones, the physical quantities
are relaxed to the initial states on a timescale varying from infinity at the damping zone edge 1.25$R_{in}$, 0.84$R_{out}$, $\theta_{in}+0.1$, and $\theta_{out}-0.1$
to a timescale of 0.1 orbit at the boundary of the simulation domain. The damping zone gradually damps waves traveling to the boundary of the simulation domain.
 Besides the wave damping zone and the long timescale, the run SM2ISOlong is different
from our main set of simulations in other ways.
We ramp the planet mass linearly for 10 orbits in run SM2ISOlong to test how our results will be affected if we insert the planet gradually in the disk. 
We also choose $\alpha=0$ in run  SM2ISOlong to confirm that the small viscosity ($\alpha=10^{-4}$) in the main set simulations will not affect
the results.}

{To verify our results, especially for the 3-D structure of spiral shocks, we have used Athena with cylindrical coordinates (run SM2ISOcyl) to 
carry out a 3-D simulation having the same disk set-up as in SM2ISO. 
This Athena simulation  (SM2ISOcyl) is different from  the Athena++ simulation (SM2ISO) in several ways.  First,  SM2ISOcyl
uses the Corner Transport Upwind Integrator which is different from the
Van-Leer Integrator used in Athena++. Second, cylindrical coordinates with uniform radial grids have been used in SM2ISOcyl while
spherical-polar coordinates with logarithmic radial grids have been used in SM2ISO.
The detailed disk set-up and the boundary conditions can be found in Zhu \etal (2014). The results will
be presented in \S 6.1. Overall, SM2ISOcyl confirms the results in SM2ISO, which greatly limits the chance
that our results are due to numerical artifacts. }

\section{The Shape of Spiral Wakes}
\begin{figure*}[ht!]
\centering
\includegraphics[trim=4cm 3cm 4cm 18cm, width=1.0\textwidth]{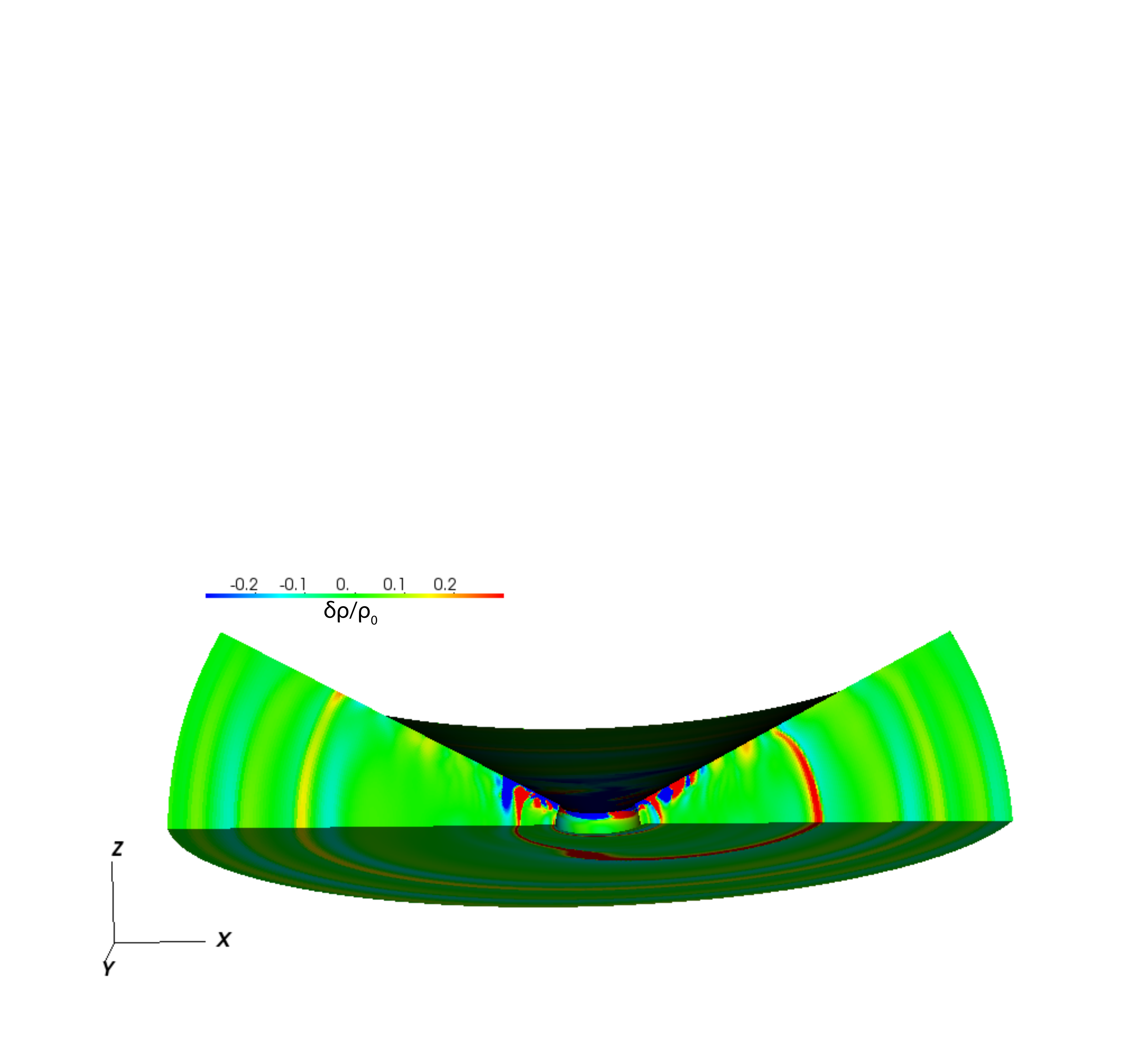} 
\vspace{0.0 cm}
\caption{Volume rendering of $\delta \rho/\rho_{0}$ for SM1ISO. The
disk has been sliced through the midplane and meridian plane to show the 3-D shock structure.
Spiral shocks have been excited by the planet, and the shocks curl towards the central star at
the disk surface. 
} \label{fig:shockvender}
\end{figure*}

Volume rendering of $\delta \rho/\rho_{0}$ in simulation SM1ISO is shown in Figure \ref{fig:shockvender}.
$\delta \rho$ is the density difference between 10 orbits and the initial condition, and
$\rho_{0}$ is the initial density at that position. Thus, $\delta \rho/\rho_{0}$ highlights the density 
perturbation (e.g. spiral shocks) in the disk. In this paper, we use ``spiral shocks'' to refer to peaks of the density wakes and 
are associated with spiral arms seen in observations.
It is apparent {in Figure \ref{fig:shockvender}} that the spiral shocks are not perpendicular to the disk midplane and they have complicated 3-D structure.
In the figure, both the inner arms inside the planet and the outer arms outside the planet curl  towards the central star at higher altitudes. 
This curled 3-D shock structure will be studied in more detail in \S 5, while in this section we focus on the shape of  the spiral wakes
in the horizontal plane.

\begin{figure*}[ht!]
\centering
\includegraphics[trim=0cm 0.cm 0cm 0cm, width=0.9\textwidth]{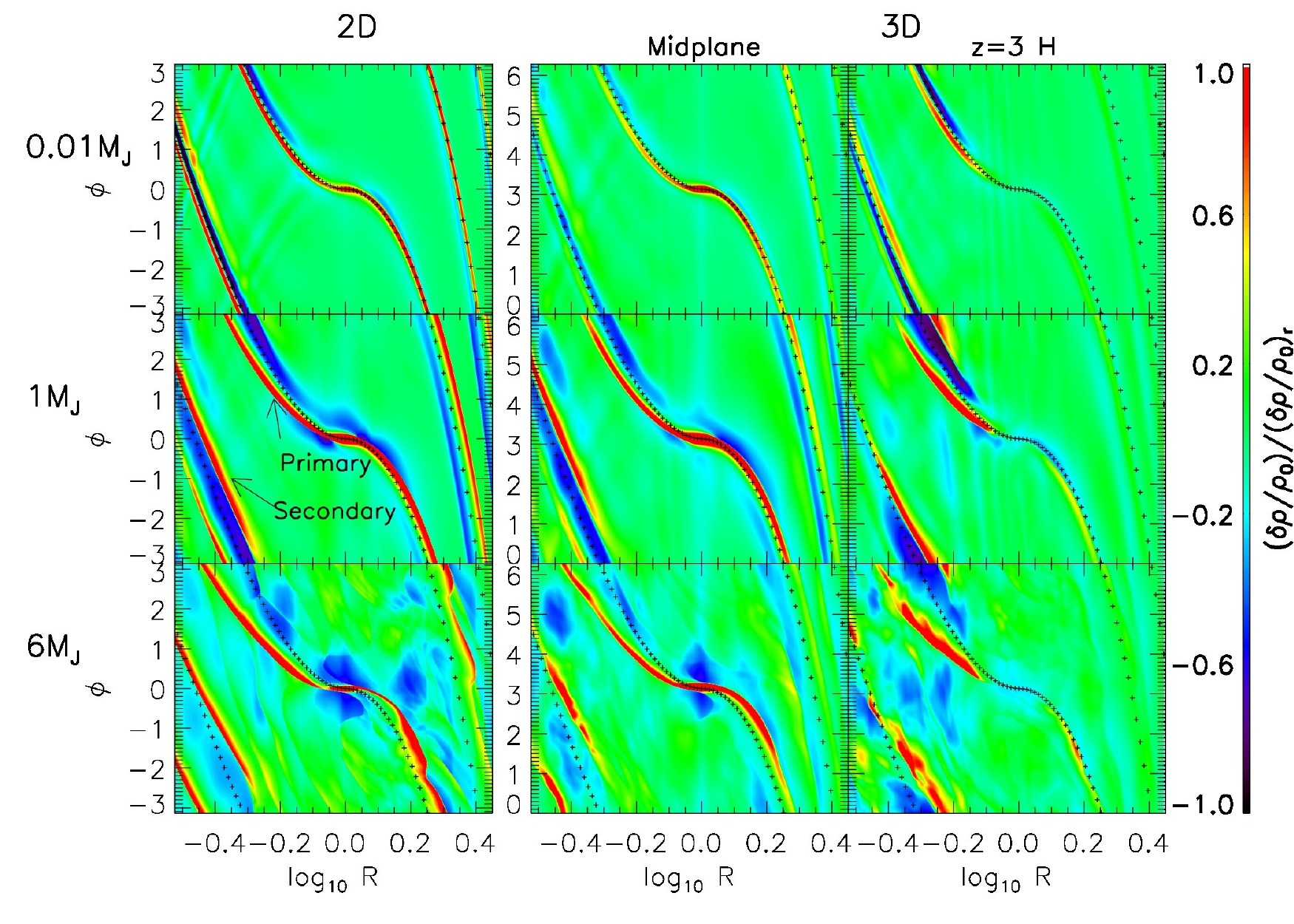} 
\vspace{0.0 cm}
\caption{$\delta \rho/\rho_{0}$ for CM1ISO, CM2ISO, CM3ISO (left panels), and  SM1ISO, SM2ISO, SM3ISO at the disk midplane (middle panels)
and $z=3 H$ (right panels). To make the spiral shocks stand out in each panel, we have scaled $\delta \rho/\rho_{0}$  to a reference value ($(\delta \rho/\rho_{0})_{r}$ ) in each panel.
$(\delta \rho/\rho_{0})_{r}$ are 0.01, 0.01, and 0.02 from left to the right panel in the first row, 0.3, 0.3, and 0.6 in the second row, and 1, 1, and 2 in the third row.
The black dotted curves are the spiral wakes from linear theory (Equation \ref{eq:linearwake} with $\alpha=3/2$ and $\eta=1/4$). When the planet is more massive, the spiral shocks have larger deviations from the prediction of linear theory. 
Due to the 3-D structure of the shocks, the inner spiral shocks become more open and the outer shocks become less open at $z=3 H$ compared with the shocks at the midplane. The color bar
is uniform but it has different scale in each plot to highlight the shock structure.
 } \label{fig:shock}
\end{figure*}

{Figure \ref{fig:shock} shows the shape of the spiral wakes in both 2-D and 3-D simulations. The x-axis is plotted in log $R$, so that the pitch angle of the spiral wake 
can be easily estimated by using its
 slope in the figure (d log $R$/d $\phi$
=tan $\beta$).}

When a very low mass planet is present in a disk, it excites density waves that are in the linear regime. The linear theory for density waves in the 2-D $R-\phi$ plane (Equation \ref{eq:linearwake}) can accurately describe the shape of the excited spiral wakes in 2-D simulations. 
This is demonstrated in the upper left panel (CM1ISO) of Figure \ref{fig:shock}
 where Equation (\ref{eq:linearwake}) with $\alpha=3/2$ and $\eta=1/4$ fits the peak of the density wakes very well\footnote{Strictly speaking, even with
 $M_{p}=0.01 M_{J}$ the excited density wakes become weak shocks at $R=0.4$ and 1.6 according to Equation (\ref{eq:eqshock}).
But the shocks are very weak and do not move away from the trajectory predicted by Equation (\ref{eq:linearwake}) significantly. }. Even at the midplane of 3-D simulations,
Equation (\ref{eq:linearwake}) still provides a good fit to the density wakes (the upper middle panel). 

However, the shape of the spiral arms at the disk surface is
affected by the 3-D structure of density wakes.  At the disk surface in 3-D simulations 
 (even in the linear regime, shown in the upper right panel of Figure \ref{fig:shock}), 
both inner and outer arms are at smaller $R$ than Equation (\ref{eq:linearwake}) 
due to the tilted shock shape in Figure \ref{fig:shockvender}. 
When these shocks
are far away from the planet, they
are more tilted towards the central star at higher altitudes. This leads to
 the inner spiral arms becoming slightly more open 
 (with a larger pitch angle) and the outer spiral arms becoming slightly less open (with a smaller pitch angle) than Equation (\ref{eq:linearwake}) would predict.

When the planet has a mass larger than $M_{th}$ (middle and bottom panels of Figure \ref{fig:shock}), 
it can launch spiral shocks immediately around the planet, and 
the shape of spiral shocks can deviate from the trajectory predicted by 
Equation (\ref{eq:linearwake}) significantly. Shocks excited by a more massive planet
deviate from linear theory more and they have larger pitch angles.
As shown in Figure \ref{fig:shock}),
spiral shocks in the 6 $M_{J}$ cases (bottom panels) 
 are more open and deviate from the prediction of linear theory (dotted curves) more than shocks in the 1 $M_{J}$ cases (middle panels). The deviation from the linear
 theory has also been seen in previous simulations, e.g., Figure 2 and 10 of de Val-Borro et al. (2006), but it has not been explored  and the physical reason for the deviation is left to be unexplained.

This deviation from linear theory shown in Figure \ref{fig:shock} is consistent with the predictions from the weakly non-linear 
density wave theory by Goodman \& Rafikov (2001) and Rafikov (2002).
In weakly non-linear theory, the spiral shock can expand in both azimuthal 
directions away from Equation (\ref{eq:linearwake}), and, at each radius, the shock density profile along the azimuthal direction  is  
N-shaped (Figure 2 in Goodman \& Rafikov 2001). 
The N-shaped shock profile expands in the azimuthal
direction at a speed which is proportional
to the normalized amplitude of the shock ($(\Sigma_{shock}-\Sigma_{0})/\Sigma_{0}$). 
Thus, the higher is the planet mass, the stronger are the shocks and these shocks 
expand  faster away from Equation (\ref{eq:linearwake}). Then, the 
spiral shock has a larger pitch angle as a result.

Similar to the 0.01 $M_{J}$ case, the inner spiral shocks in 1 and 6 $M_{J}$ cases are even more open at $z=3H$ than at the midplane, while the outer arms
become less open at the disk surface. 
For outer spiral arms, this 3-D effect compensates the increased pitch angel due to 
the shock expansion, and
coincidently the outer arms almost overlap with the prediction from linear theory. 

Another important feature shown in Figure \ref{fig:shock} is that, besides the primary inner arm which originates from the planet,  a secondary inner spiral arm appears with some azimuthal shift from the primary arm. {For some cases (e.g. CM2ISO), we even see a tertiary arm at the very inner disk.}
The secondary spiral arm has also been seen in previous simulations having massive planets, e.g. Figure 2 in Kley (1999) and Figure 10 of de Val-Borro et al. (2006). However, it has hardly been explored in earlier simulations. Figure \ref{fig:shock} shows that, even with a
very low mass planet (0.01 $M_{J}$, the upper panels of Figure \ref{fig:shock}), another density peak (the secondary arm) emerges close to the primary inner 
arm with low density region (the rarefaction wave of the primary arm)
in between. After the secondary inner arm appears close to the primary arm, it can become shock during the propagation and later
it will become N-shaped which is similar to the primary arm. Then in some cases, a tertiary inner arm appears at the rarefaction wave part of the secondary arm.
For 0.01 $M_{J}$ cases, the primary and secondary inner arms are separated by $\delta \phi\sim1$ 
at $R=0.3$. When the planet gets more massive, the secondary inner arm is excited earlier  and the separation between the primary and secondary arm increases. In 1 $M_{J}$ cases, the two inner arms are roughly separated by $\delta \phi\sim2$, and in  6 $M_{J}$ cases, the two arms are roughly separated by $\delta \phi\sim3$. This has an important application that we can
use the separation between two arms to estimate the mass of the embedded planet. 
Similar to the primary inner arm, the secondary inner arm is also 
 stronger at the disk surface than at the disk midplane. 
For outer arms, the secondary arm  appears in disks that have massive planets embedded, but the secondary outer arm is less apparent than the primary outer arm. More discussions on the 3-D structure of secondary arms will be presented in \S 5.

\begin{figure*}[ht!]
\centering
\includegraphics[trim=0cm 0.6cm 0cm 0cm, width=0.8\textwidth]{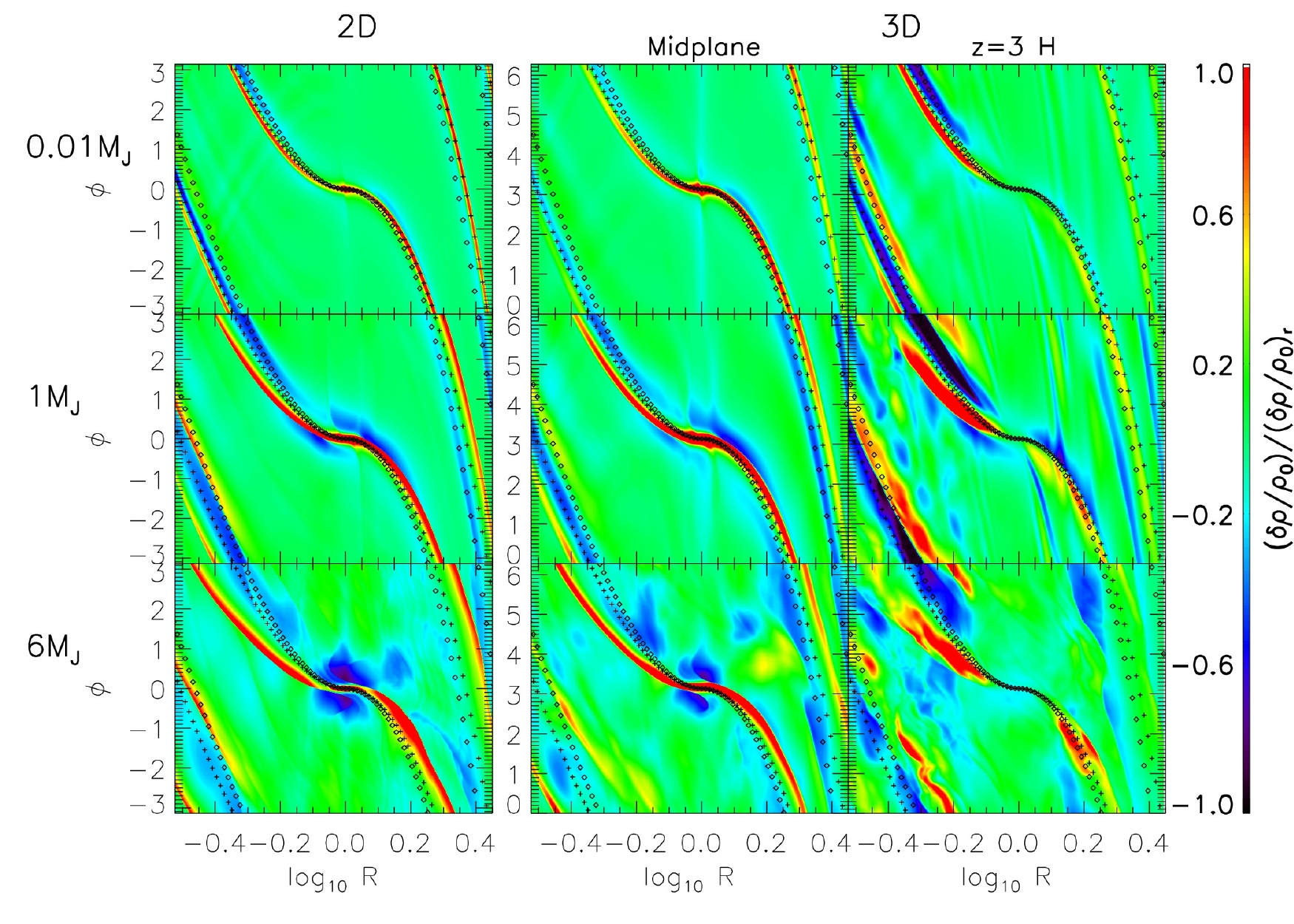} 
\vspace{0.0 cm}
\caption{ The same as Figure \ref{fig:shock} but for CM1T1, CM2T1, CM3T1 (left panels), and  SM1T1, SM2T1, SM3T1. 
$(\delta \rho/\rho_{0})_{r}$ are 0.005, 0.005, and 0.01 from left to the right panel in the first row, 0.3, 0.3, and 0.6 in the second row, and 0.8, 0.8, and 2.4 in the third row.
The black squared dots represent the linear theory using isothermal sound speed while the black plus sign dots use the adiabatic sound speed. } \label{fig:shocktc1}
\end{figure*}

Spiral wakes/shocks are slightly more open in a disk whose EoS is not isothermal (Figure \ref{fig:shocktc1}). This is because
density waves propagate slightly faster in a fluid with a non-isothermal EoS than in a fluid at the same temperature with the isothermal EoS (e.g. Goodman \& Rafikov 2001). 
In Figure \ref{fig:shocktc1}, even with a moderate cooling rate ($T_{cool}=1$), 
the spiral wakes excited by a low mass planet (0.01 
$M_{J}$) can only be fitted by Equation (\ref{eq:linearwake}) using a larger disk scale height that is calculated
with the adiabatic sound speed instead of the isothermal sound speed
 ($h_{p}$ in Equation \ref{eq:linearwake} is thus $c_{s,adi}/R\Omega=\sqrt{\gamma}c_{s,iso}/R\Omega$). 
{It is a little bit surprising that the spiral shape in disks with $T_{cool}=1$ follows the spiral shape in adiabatic disks. 
We think this is due to the short timescale for the flow to move across the spiral wake.  The spiral wake has a typical width smaller than the disk scale height, while
the background flow moves across the wake at nearly Keplerian speed. Thus, when the fluid travels in and out of the wave/shock, its response time is much smaller than the orbital time
and thus behave adiabatically.}

Other aspects of the
 spiral shocks in non-isothermal cases are similar to the isothermal cases, e.g., a higher mass planet excites a more open spiral shock, and the inner spiral shocks become slightly
more open at higher altitudes. 

\begin{figure*}[ht!]
\centering
\includegraphics[trim=0cm 0.2cm 0cm 0.2cm, width=0.9\textwidth]{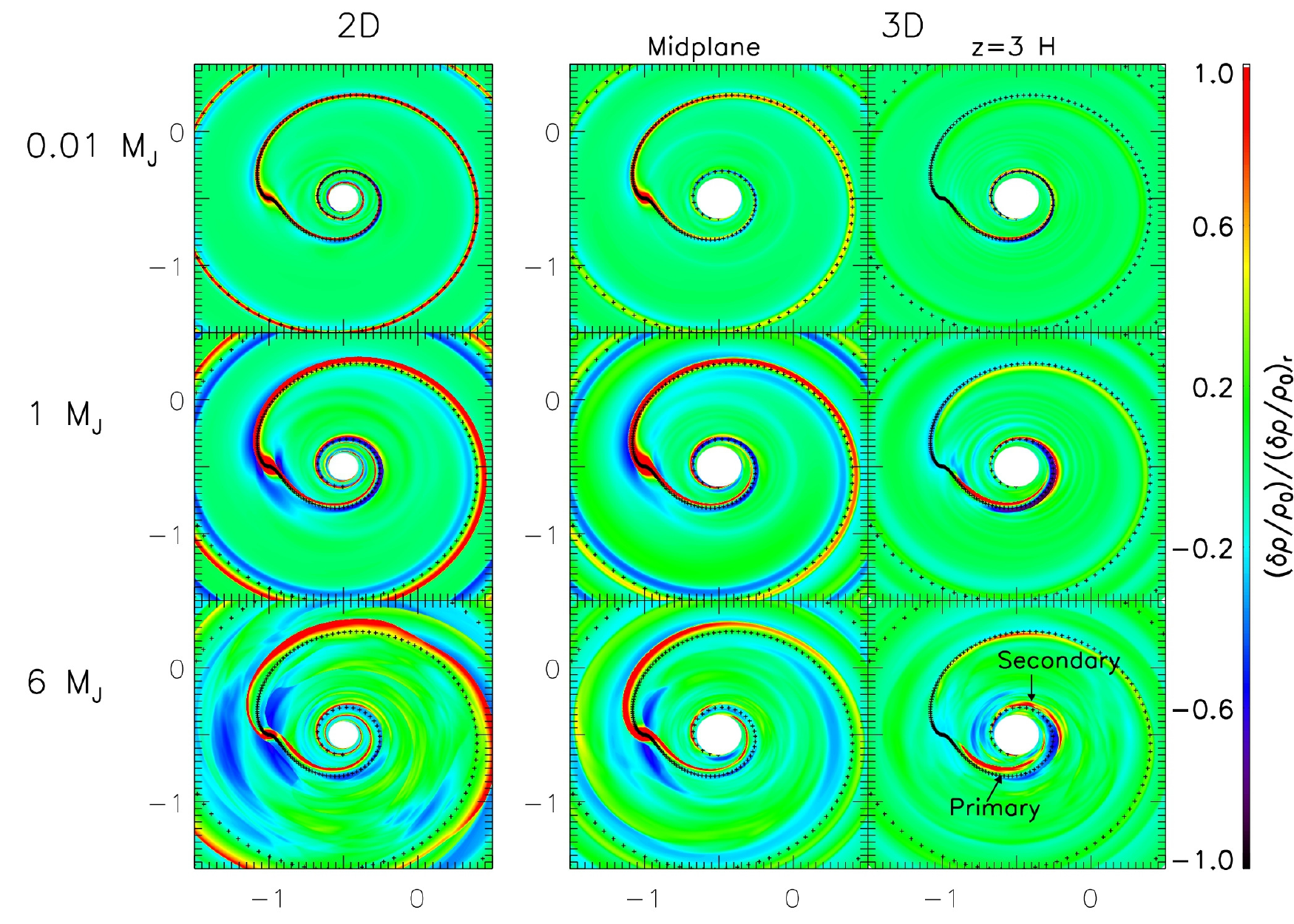} 
\vspace{0.0 cm}
\caption{The same as Figure \ref{fig:shock} but in Cartesian coordinates.  Please note that $(\delta \rho/\rho_{0})_{r}$ 
is different in each panel to make the spiral arms stand out. } \label{fig:shockcarg}
\end{figure*}

 To illustrate the shape of the spiral shocks in the physical space, we plot the relative density perturbation in Cartesian coordinates in Figure \ref{fig:shockcarg}. Clearly, the more massive the planet is, the more the spiral shocks deviate from linear theory. Two well separated 
inner arms are also apparent when the planet mass is large, and the separation between these
two arms is larger when the planet is more massive (comparing the middle and bottom panels in Figure \ref{fig:shockcarg}). 

\begin{figure*}[ht!]
\centering
\includegraphics[trim=0cm 0.8cm 0cm 0cm, width=1.0\textwidth]{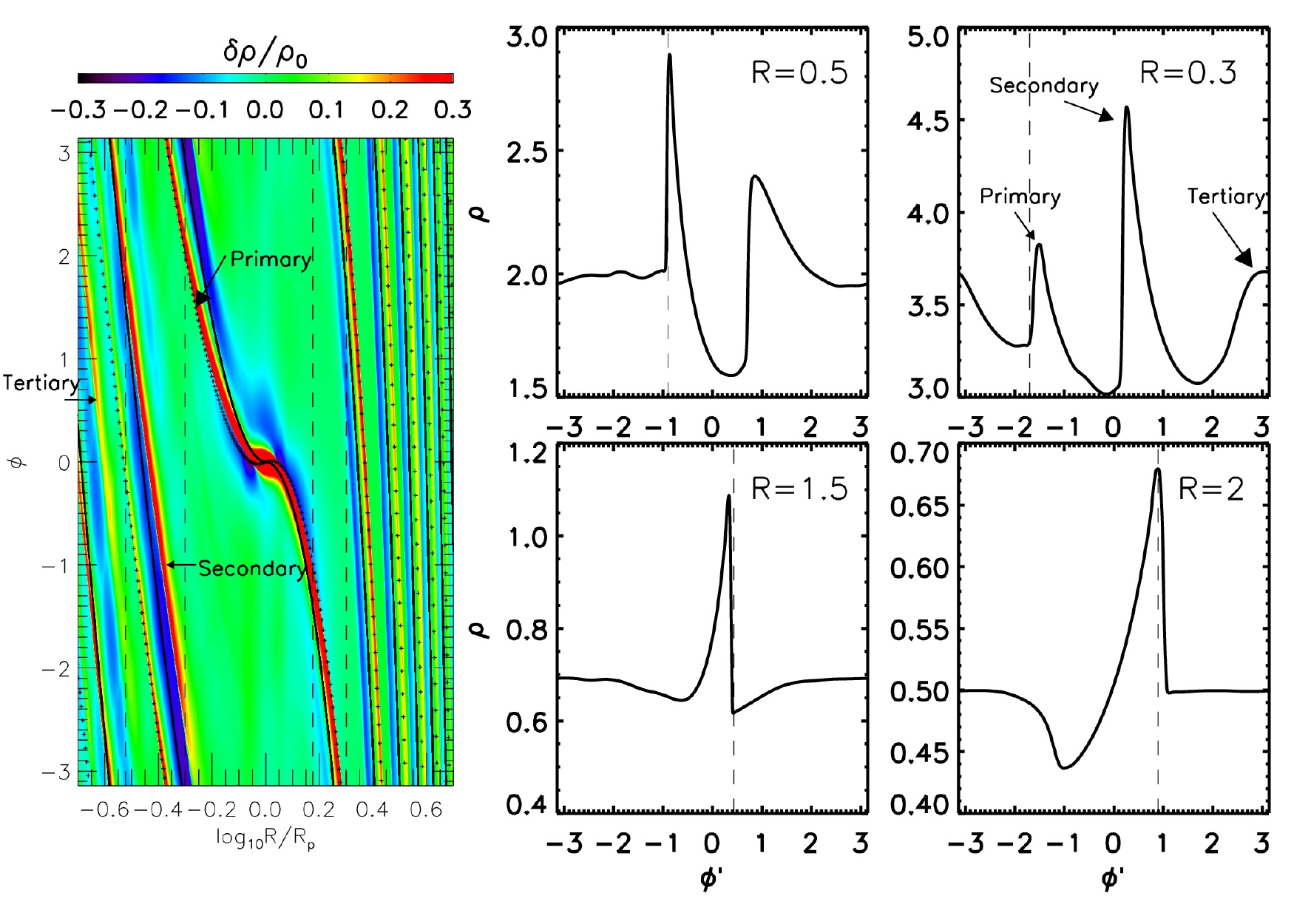} 
\vspace{0.0 cm}
\caption{$\delta \rho/\rho_{0}$ for CM2ISO (left color panel) and density cuts across the $\phi$ direction at $R=$0.5, 0.3, 1.5, and 2.0.  
The solid black curve in the left color panel labels the shock position from linear theory (Equation \ref{eq:linearwake}). 
In the density cut plots, the density profile has
been shifted so that the shock position from linear theory is at $\phi'=0$. The dashed lines in $R$=0.5 and 1.5 plots
label the shock fronts, while the dashed lines in $R$=0.3 and 2 plots label the predicted shock fronts from weakly nonlinear theory. In the left color contour panel, the predicted shock front from nonlinear theory is labeled as the dotted curve. } \label{fig:shockcut}
\end{figure*}

To see how successful the weakly non-linear density wave theory of Goodman \& Rafikov (2001) and Rafikov (2002) fits the shape of the shocks in simulations, 
we plot in Figure \ref{fig:shockcut} the density contour and density profiles along the azimuthal direction at $R=0.3$, 0.5, 1.5 and 2 for run CM2ISO. In the density contour panel,
we can clearly see that the secondary arm appears at the edge of the low density rarefaction wave region of the primary arm. 
{After the secondary arm propagates for a short distance,
a tertiary arm appears close to the low density rarefaction wave region of the secondary arm.} In the panels showing the density profiles, we
follow Goodman \& Rafikov (2001) and Rafikov (2002) to shift the density profiles  so that $\phi'=0$ corresponds to the
wake position from linear theory (black curve in Figure \ref{fig:shockcut} or Equation \ref{eq:linearwake}). 
The density profiles clearly show that the shocks are N-shaped. A rarefaction wave follows the shock front, and
$\delta \rho$ at the rarefaction wave region can  be negative before it merges to the background flow. The shock fronts 
deviate from $\phi'=0$, and due to the shock expansion, the deviation is larger when the shock is further away from the planet. 
Under the shearing-sheet approximation, 
 the amplitude and width of the N-shaped
shock scale as $|R-R_{p}|^{-5/4}$ and $|R-R_{p}|^{5/4}$ at $|R-R_{p}|\gg$0 based on the weakly 
non-linear density wave theory of Goodman \& Rafikov (2001). In a global disk spanning
a large range of radii, 
 the amplitude and width of the N-shaped shock scale as $t^{-1/2}$ and $t^{1/2}$, where $t$ is given
 in Equation 43 of Rafikov (2002). For our disk parameters, we have
 \begin{equation}
 t\propto \left|\int_{1}^{R/R_{p}}|s^{3/2}-1|^{3/2}s^{-13/8}ds\right|\,.\label{eq:tf}
 \end{equation}
 Since the middle point of the N-shaped shock is at $\phi'\sim$0 (Rafikov 2002),
we expect that in our global simulations the azimuthal distance between the shock front and the path 
predicted by Equation \ref{eq:linearwake} ($\phi'$=0) should also scales as  $t^{1/2}$. 

To test this prediction, we have measured
the shock positions at $R=0.5$ and 1.5 in Figure \ref{fig:shockcut}, which are $\phi'=-0.9$ and 0.43 respectively. These two positions are labeled
as the dashed lines in $R=0.5$ and 1.5 panels. Then we calculate the shock positions at $R=0.3$ and 2 to be -1.7 and 0.9, using 
their positions at $R=0.5$ and 1.5 together with the scaling relationship $t^{1/2}$ where $t$ at $R$=0.3, 0.5, 1.5, and 2 are calculated from Equation \ref{eq:tf}. These predicted shock positions are
labeled as the dashed lines in  $R=0.3$ and 2 panels. We can see that they agree with the actual shock positions in the simulation
very well. This confirms that the shock positions are determined by the non-linear expansion of spiral shocks. Using the same approach,
we have calculated the non-linear shock position at every radius for $R<1$ and $R>1$  with
the normalization based on shock positions at $R=0.5$ and 1.5. This new predicted shock shape is plotted as the 
dotted curve in the left panel of Figure \ref{fig:shockcut}. Despite some offset at $R$ close to the planet, which is expected since the simple relationship $t^{-1/2}$ holds only when $|R-R_{p}|\gg$0,
a good agreement has been achieved between the non-linear density
wave theory and the simulations. Note that in this comparison, we did not calculate the shock
strength directly from non-linear theory and compare its amplitude with simulations, 
instead we use the scaling relationship to verify
the propagation of the shock. In future, direct comparison is desired when we have a more complete non-linear theory which can calculate the shock excitation directly. 

Although the primary arm can be fitted by the weakly non-linear density wave theory,
 the excitation of the secondary (or even tertiary) arm  still lacks a good theoretical explanation. It may be related
 to the low $m$ mode (e.g. $m=2$, $3$, similar to disks in binary systems) or some non-linear wave coupling. 
Figure \ref{fig:shockcut}  suggests that a secondary spiral arm is excited at the other end of the N-shaped primary shock (the $R=0.5$ panel). After it is excited, it steepens to shocks and becomes another N-shaped shock later (the $R=0.3$ panel). Its shock front can even travel into the rarefaction
wave of the primary arm (e.g. in the $R=0.3$ panel, the secondary shock is almost at $\phi'$=0 where the rarefaction wave of the primary arm should reside.). 
 Unlike the primary arm which already dissipates
when it travels inward  from $R=0.5$ to 0.3, this secondary arm is excited later and becomes stronger from $R=0.5$ to 0.3. 
 At $R=0.3$, the secondary arm is even stronger than the primary arm. 
By comparing $R=0.5$ and $R=0.3$ panels, we also notice that the secondary arm 
 almost keeps the same azimuthal separation  with the primary arm ($\Delta \phi'\sim$1.7) during its propagation.
{Furthermore, at $R=0.3$, a tertiary arm starts to appear at the other end of the N-shaped secondary arm. }  
  
\begin{figure}[ht!]
\centering
\includegraphics[trim=0cm 0.8cm 0cm 0cm, width=0.5\textwidth]{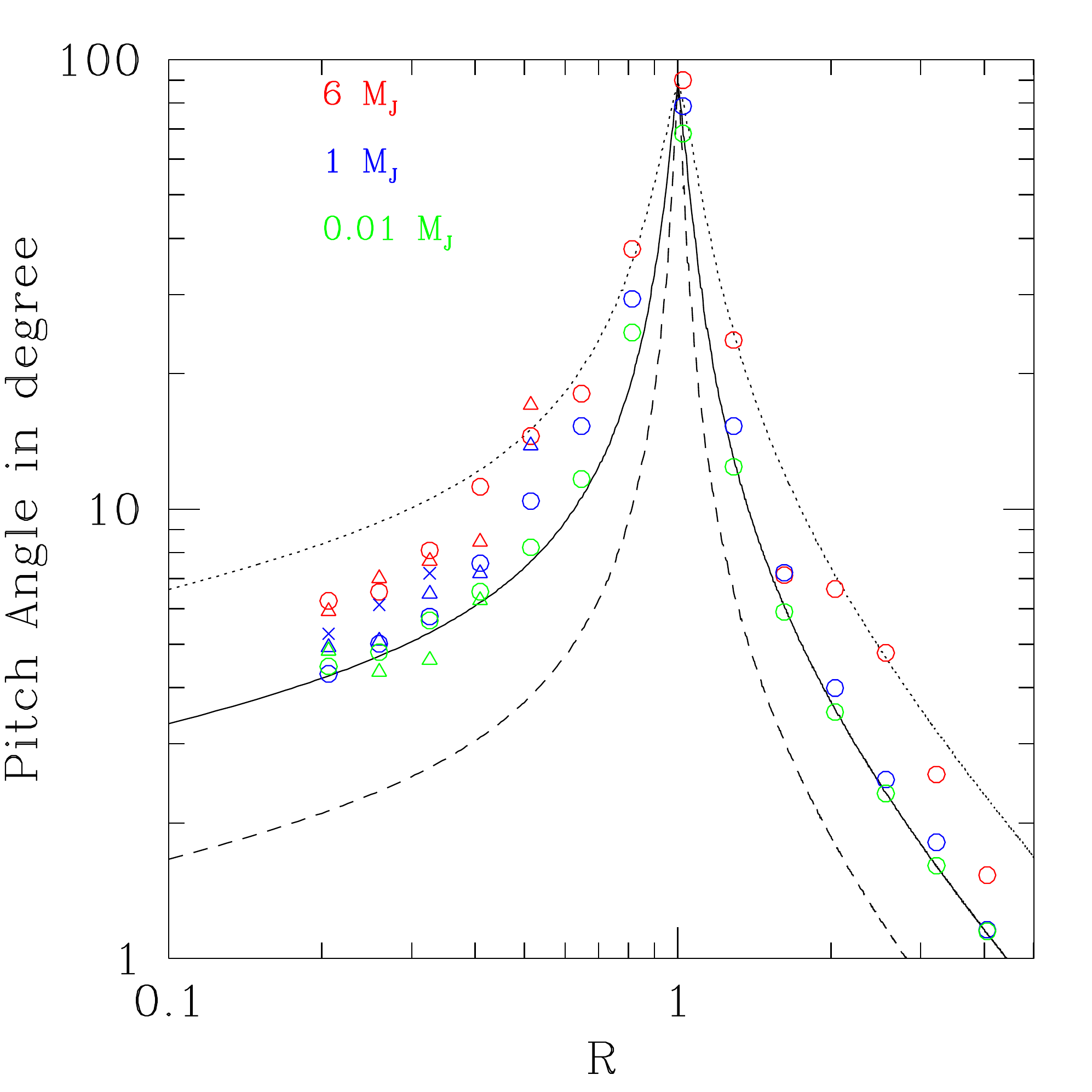} 
\vspace{0.0 cm}
\caption{The pitch angle based on the linear theory (Equation \ref{eq:linearwake} with $\alpha=3/2$ and $\eta=1/4$) assuming  $h_{p}$=0.1 (solid curve), 
0.2 (dotted curve), and 0.05 (dashed curve) compared with those measured in numerical simulations. The red, blue, and green dots are measured from 
CM3ISO, CM2ISO, and CM1ISO respectively.  The circles, triangles, and crosses are the pitch angle of the primary, secondary and tertiary arms. 
} \label{fig:pitch}
\end{figure}

{To summarize the results in this section, Figure \ref{fig:pitch} shows the pitch angle from the linear theory and those measured
in numerical simulations. When the planet mass is low (e.g. CM1ISO, the green dots), the measured pitch angle agrees with the linear theory. 
When the planet mass increases, the pitch angle also increases. For the 6 $M_{J}$ case, the measured pitch angle of the spiral wake in the $h_{p}=0.1$ disk
is close to the pitch angle predicted in a much thicker disk ($h_{p}=0.2$) using the linear theory. The secondary and even the tertiary arms have
similar pitch angles as the primary arms. Thus, if we know the disk thermal structure very well, we can use the deviation of the measured 
pitch angle from the linear theory to estimate the embedded planet mass. }
  
\section{The 3-D Structure of Spiral Wakes}

\begin{figure}[ht!]
\centering
\includegraphics[trim=0cm 0.8cm 0cm 0cm, width=0.5\textwidth]{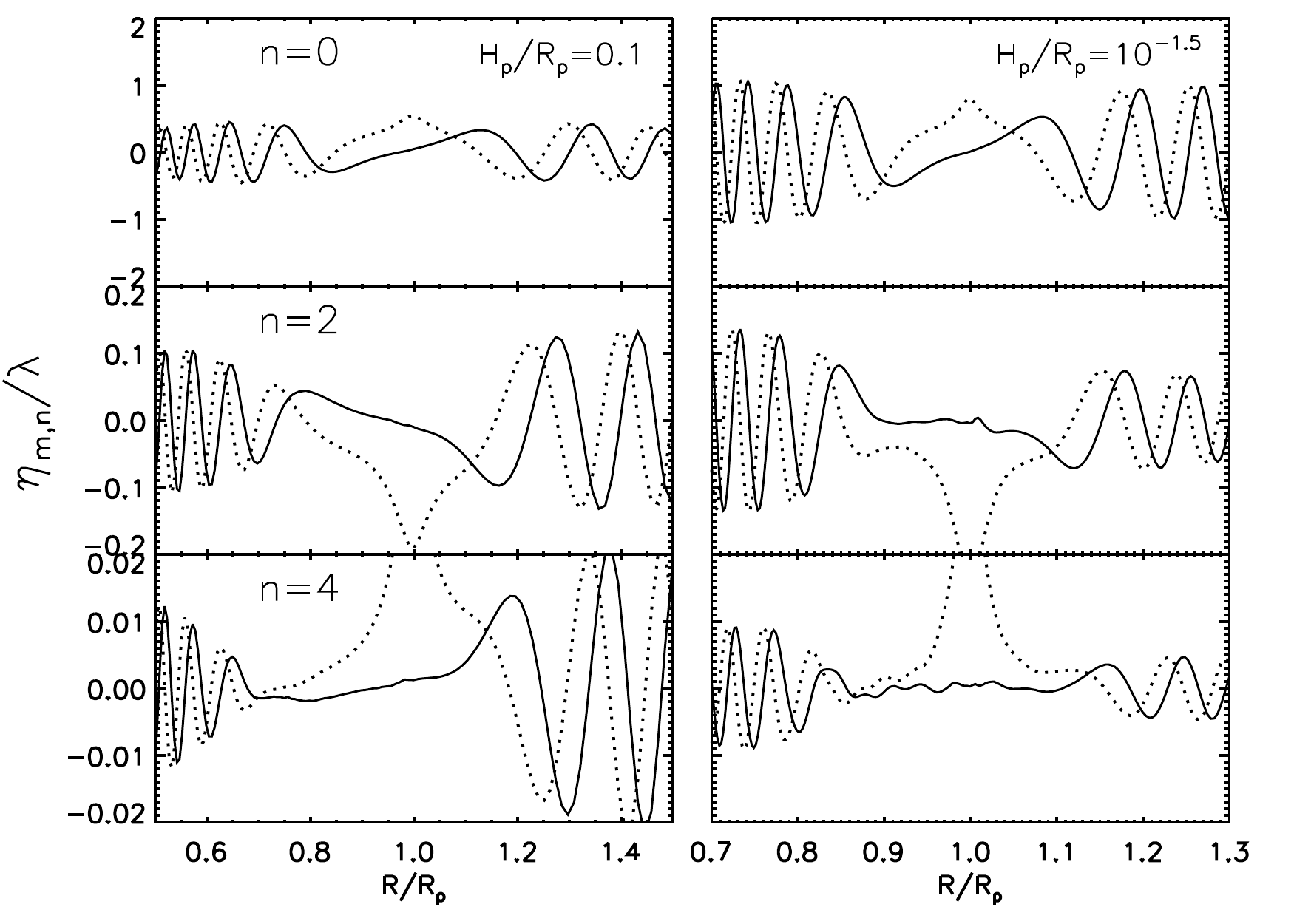} 
\vspace{0.0 cm}
\caption{Waves excited on a 3-D disk by a low mass planet in run SM1ISO (left panels) and STHIN (right panels). Density perturbations
with $n$=0,2, and 4 Hermite components for the m=10 Fourier mode are displayed. The real (imaginary) part of $\eta$ is shown with
a dotted (solid) curve. Y-scales are different for different $n$ modes.} \label{fig:m10}
\end{figure}

Since near-IR scattered light observations only probe the shape of the disk surface,
3-D structure of spiral shocks can affect the observational signatures of these spiral shocks.
Intuitively, we would expect that the spiral shocks have complicated 3-D structure. 
First, the wave excitation must have 3-D structure since, at the same $R$ in the disk, 
the distance between
the planet and the disk surface is larger than the distance at the midplane, and the force
is thus weaker at the disk surface. Second, the wave propagation may have 3-D structure
considering the disk becomes thinner at smaller $R$. Waves/shocks are more converged
when they propagate inwards. They can also channel to the disk surface (Lubow \& Ogilvie 1998, Bate et al. 2002, Lee 
\& Gu 2015),
and, during their propagation from the high density region (e.g., the midplane) to the
low density region (e.g., the  disk surface), the amplitudes of perturbations have to increase
 to conserve the wave action. Since the amplitudes of perturbations determine when the waves 
 will break into shocks, the dissipation can also be quite different between the surface and the midplane. All these effects
can contribute to the 3-D structure of spiral waves/shocks.

Due to these complicated effects, it is difficult to develop an analytic theory to study the planet-induced 3-D shock structure, and we rely on numerical simulations to study
such structure. However, before delving into the highly nonlinear shock regime, 
 we can use the linear theory developed
in Tanaka, Takeuchi \& Ward (2002) to estimate the 3-D effect of the density waves. Following their theory for locally isothermal disks,
the structure of the waves in the $z$ direction can be studied with Hermite polynomials ($H_{n}(Z)$). 
We first expand perturbed quantities ($\eta$) from our simulations into Fourier series 
\begin{equation}
\eta=\underset{m}{\sum} {\rm Re} \left[\eta_{m}e^{im(\phi-\Omega_{p}t)}\right]\,,
\end{equation}
where the Fourier components $\eta_{m}$ are complex functions of $R$ and $z$. Then, $\eta_{m}$ can be further expanded
with Hermite polynomials in the $z$ direction,
\begin{equation}
\eta=\overset{\infty }{\underset{m=0}{\sum} }\overset{\infty }{\underset{n=0}{\sum} }{\rm Re} \left[\eta_{m,n}H_{n}(Z)e^{im(\phi-\Omega_{p}t)}\right]\,,
\end{equation}
where $Z$ is the normalized height as $Z=z/H(R)$, and the first three Hermite polynomials are
\begin{equation}
H_{0}(Z)=1\,,\quad H_{1}(Z)=Z\,,\quad H_{2}(Z)=Z^2-1\,.
\end{equation}
By using the normal orthogonal relation between $H_{n}$, we have
\begin{equation}
\eta_{m,n}=\frac{1}{\sqrt{2\pi}n!}\int_{-\infty}^{\infty}e^{-Z^2/2}H_{n}(Z)\eta_{m}dZ\,.\label{eq:mn}
\end{equation}
We can use $\eta_{m,n}$ at different $n$ to estimate 
the relative importance of different Hermite components. 
To compare with Figure 2 in Tanaka \etal (2002), we show $m=10$, $n=0, 2, 4$ Fourier-Hermite components for
the perturbed density  ($\delta \rho/\rho_{0}$)\footnote{ Since the disk is isothermal locally, the density perturbation is quite similar to the enthalpy perturbation, and can be compared with  Figure 2 in Tanaka \etal (2012).} in 
Figure \ref{fig:m10}.
With the same parameters, the right panel of Figure \ref{fig:m10} is very similar to Figure 2 in Tanaka \etal (2012). 
By comparing the left and right panels in Figure \ref{fig:m10}, we can see that although higher-order Hermite components have similar strength between thin
and thick disks, n=0 Hermite component is much weaker in a thick disk, suggesting that the wakes in a
 thick disk have more significant 3-D structures than wakes in a thin disk.

Figure \ref{fig:m10} suggests that higher-order vertical components can dominate the disk structure at 
the atmosphere. Although it
shows that the $n=4$ component ($\eta_{10,4}$) is 10 times weaker than the $n=2$ component ($\eta_{10,2}$),
and the $n=2$ component is 10 times weaker than the $n=0$ component (which led Tanaka \etal 2002 to conclude that most of the angular momentum excited by the planet will be carried by
two dimensional free waves), the base function (Hermite polynomials) at $z=3H$ gets $\sim$ 10 times larger sequentially from $H_{0}$ to $H_{2}$ and to $H_{4}$. Thus, $\eta_{10,2}H_{2}$ and $\eta_{10,4}H_{4}$ are still
comparable with $\eta_{10,0}H_{0}$.
The density structure at the disk atmosphere can be significantly affected by higher-order
vertical modes. 

\begin{figure}[ht!]
\centering
\includegraphics[trim=0cm 0.8cm 0cm 0cm, width=0.5\textwidth]{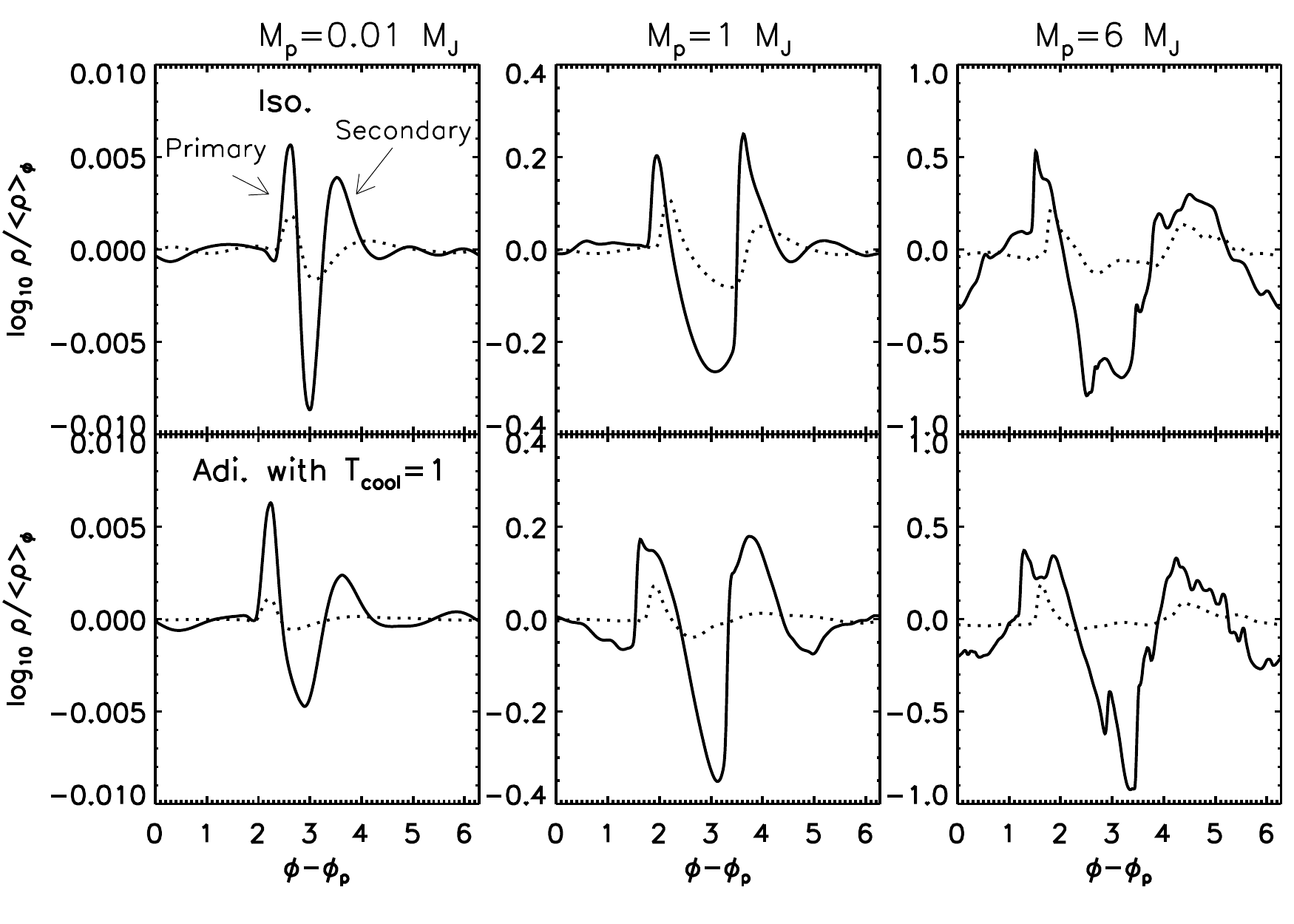} 
\vspace{0.0 cm}
\caption{At the disk radius of $R=0.5$, density profiles along the azimuthal direction at the disk midplane (dotted curves) and  $z=3 H$ (solid curves). 
$\langle\rho\rangle_{\phi}$ is the density averaged over the azimuthal direction.
Simulations with different planet masses (different columns) and equations of state (upper panels: isothermal, bottom panels: adiabatic with $T_{cool}=1$) 
have been shown.} \label{fig:onedr0p5}
\end{figure}

Although the modal analysis is useful to verify numerical simulations and can be suggestive
on the relative amplitudes of various modes, it is the
 3-D structure in the real space that determines the observational signatures of waves/shocks.

By studying the shock structure in real space, we first find that the 3-D shock structure is dramatically different between inner
and outer arms. 
 For the inner arms, the density perturbation is much larger at the disk surface than 
at the disk midplane. The 3-D structure of the inner spiral arms at $R=0.5$ is shown in Figure \ref{fig:onedr0p5}. At  $R=0.5$ and $z=3 H$ (solid curves), the differences between the maximum
and minimum density 
in the logarithmic scale are 0.015, 0.4, and 1.3 for M1, M2, and M3 cases respectively, in comparison with 0.004, 0.2, and 0.4 at the disk midplane (dotted curves).
At the same radius ($R=0.5$), the position of the wakes in non-isothermal disks (lower panels) are at smaller $\phi-\phi_{p}$ compared with
those in isothermal disks (upper panels). This is because the wakes are more open in non-isothermal disks, as discussed in \S 4, .

\begin{figure}[ht!]
\centering
\includegraphics[trim=0cm 0.8cm 0cm 0cm, width=0.5\textwidth]{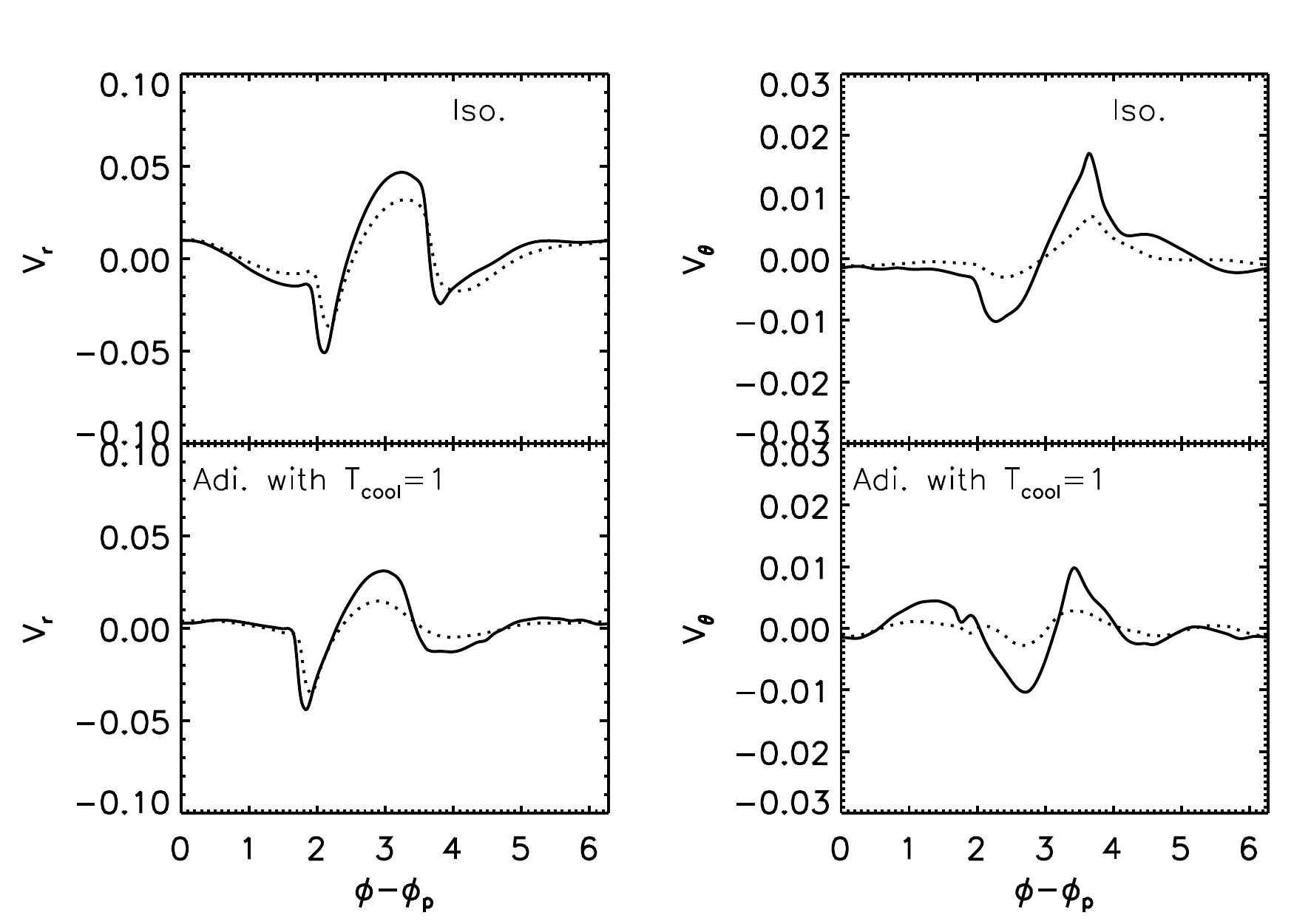} 
\vspace{0.0 cm}
\caption{$v_{r}$ (left panels) and $v_{\theta}$ (right panels) at $z=1 H$ (dotted curves) and $2 H$ (solid curves) for SM1ISO (upper panels)
and SM1T1 (bottom panels). $v_{\theta}$ is positive when the motion is towards the disk surface.} \label{fig:vyr0p5}
\end{figure}

The secondary inner spiral arms/shocks are also more prominent at the disk surface than at the disk midplane. At the disk midplane,
the secondary arms have lower amplitudes compared with the primary arms (dotted curves in Figure \ref{fig:onedr0p5}), while at $z=3 H$, the secondary arms  have almost the same amplitudes
as the primary arms (solid curves). The large amplitude of primary and secondary inner arms 
at the disk surface is due to the corrugated motion in the $v_{\theta}$ direction. 
In Figure \ref{fig:onedr0p5} which is in the corotating frame with the planet, the disk material flows in the direction from the left side to the right side of the figure. Before meeting with the shock, the disk is in vertical hydrostatic equilibrium with the background density 
and $v_{\theta}=0$ (Figure \ref{fig:vyr0p5}). 
After the shock, the disk material loses angular momentum and moves inwards with $v_{r}<0$ (Figure \ref{fig:vyr0p5}). 
At the same time, $v_{\theta}$ also becomes
negative, compressing the disk material at the midplane. This downward motion decreases
the density of the rarefaction wave at $z=3 H$.  Before meeting the secondary shock, $v_{\theta}$ starts to increase and becomes positive, 
  leading to a higher density at the disk surface. At the secondary shock, $v_{\theta}$ reaches the maximum positive
 velocity and leads to the highest density at the disk surface for the secondary shock in Figure \ref{fig:onedr0p5}. This  corrugated motion, first negative 
and then positive
$v_{\theta}$, leads to an enhanced contrast between the spiral shock and the rarefaction wave after the shock.

\begin{figure}[ht!]
\centering
\includegraphics[trim=0cm 0.8cm 0cm 0cm, width=0.5\textwidth]{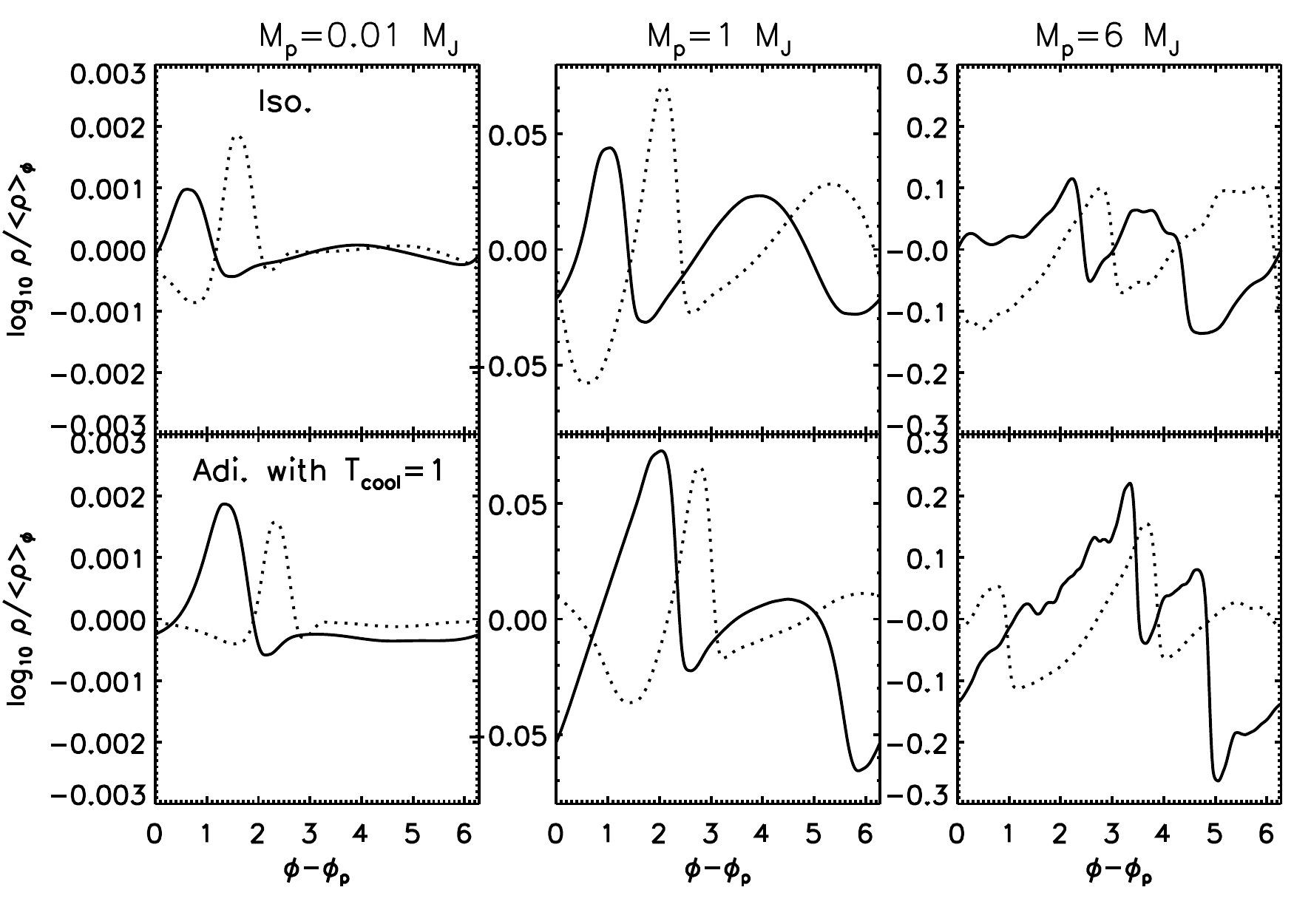} 
\vspace{0.0 cm}
\caption{Similar to Figure \ref{fig:onedr0p5}, but at $R=2$.} \label{fig:onedr2}
\end{figure}

\begin{figure}[ht!]
\centering
\includegraphics[trim=0cm 0.8cm 0cm 0cm, width=0.5\textwidth]{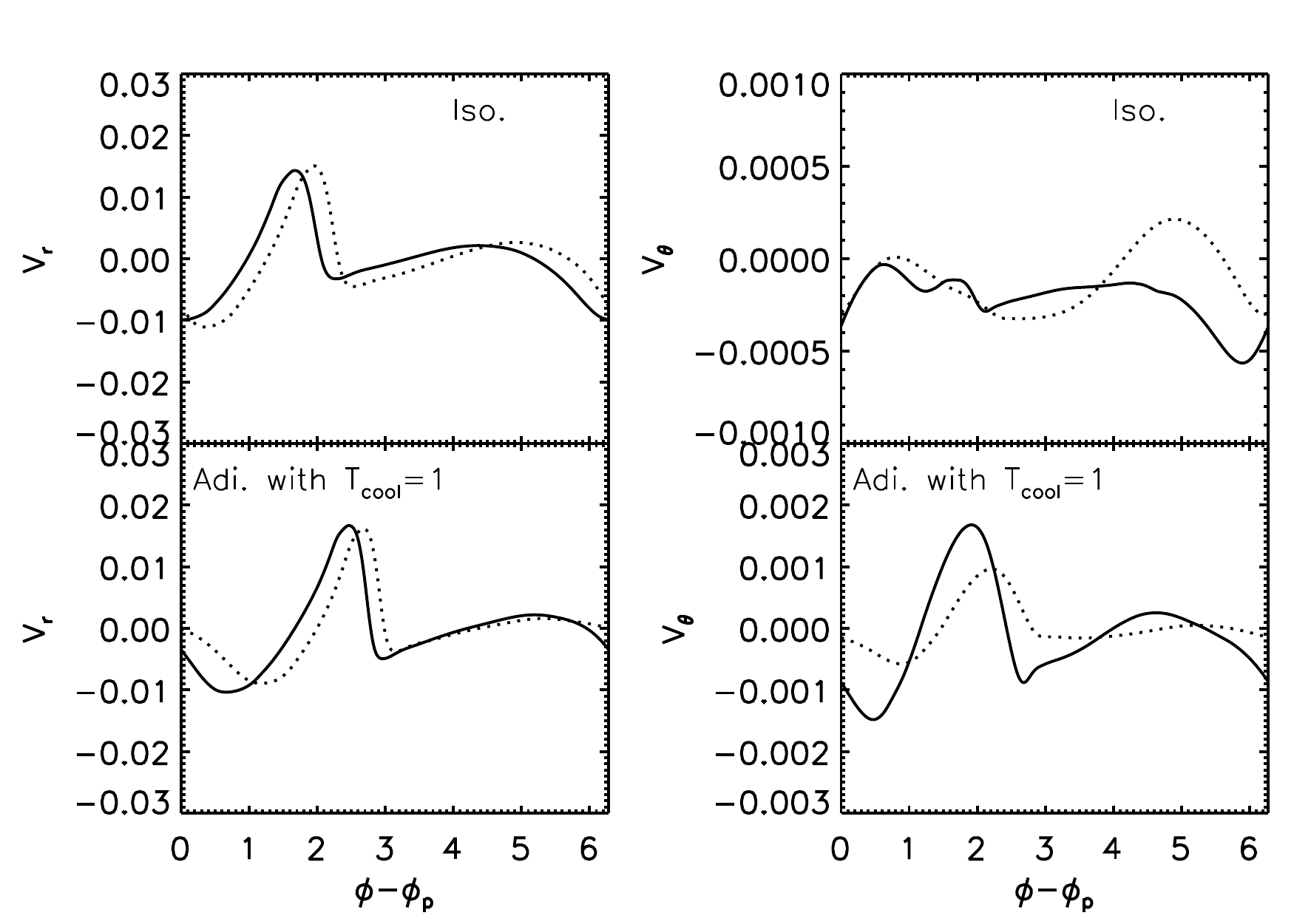} 
\vspace{0.0 cm}
\caption{Similar to Figure \ref{fig:vyr0p5}, but at $R=2$.} \label{fig:vyr2}
\end{figure}

On the other hand, the density perturbation of outer spiral arms is similar between the disk surface and the disk midplane, 
especially for isothermal disks, as shown in Figure \ref{fig:onedr2}. At  $R=2$, regardless of height,
the differences between the maximum
and minimum density 
in the logarithmic scale are both 0.002 for SM1ISO, 0.1 for SM2ISO , and 0.3 for SM3ISO.
This lack of vertical variation is also reflected in Figure \ref{fig:vyr2} where $v_{\theta}$ is very small compared with $v_{r}$ ($v_{\theta}$ is almost two orders of magnitude smaller than $v_{r}$). 
Thus, the density structure of the outer spiral is mainly determined by $v_{r}$ and $v_{\phi}$ in the horizontal plane in stead of the corrugated  motion in the $\theta$ direction.   

For non-isothermal runs (bottom panels in Figure \ref{fig:onedr2}), the density perturbation of the outer spiral arms 
at the disk surface is slightly higher than the perturbation at the disk midplane. 
Disk material flows from the right hand side of the figure to the left hand side in Figure \ref{fig:onedr2} and \ref{fig:vyr2}. When it 
meets the shock, it develops a $v_{\theta}$ towards the disk surface, which enhances the density at the disk surface. Although it is tempting to contribute
such  difference between isothermal
and non-isothermal  runs to the nonlinear hydraulic jumps (shock bores) (Boley \& Durisen 2006),  such disk structure also 
appears even in the linear regime for the 0.01 $M_{J}$ case, implying that it may be a linear effect and
 related to the eigenfunctions of the 3-D waves excited by the planet.

\begin{figure*}[ht!]
\centering
\includegraphics[trim=0cm 1.8cm 0cm 1.2cm, width=1.0\textwidth]{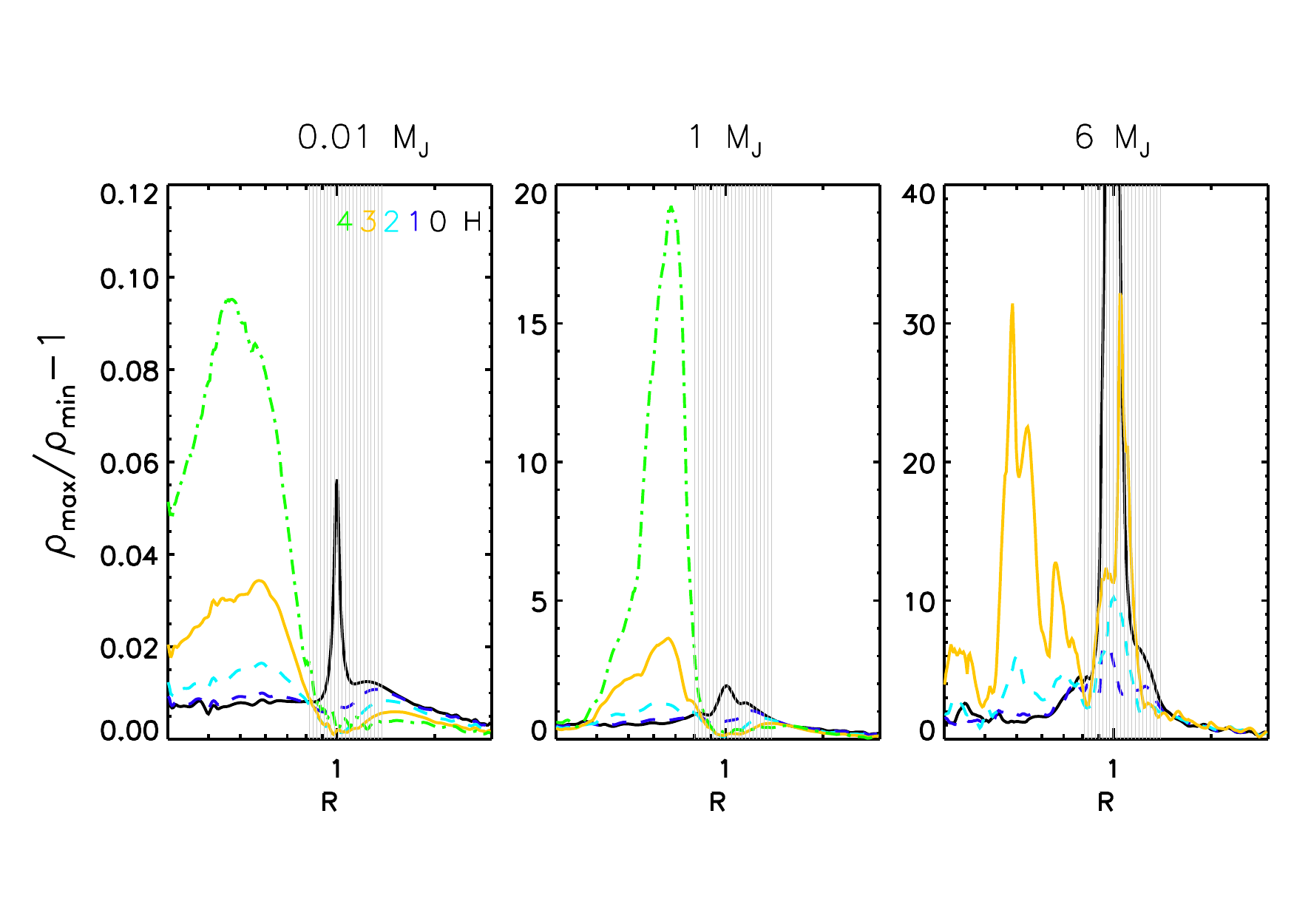} 
\vspace{0.0 cm}
\caption{Relative density perturbations at different heights for SM1ISO, SM2ISO, SM3ISO. $\rho_{max}$ and $\rho_{min}$
are the maximum and minimum density along the circle in the azimuthal direction at  given $R$ and $z$. The black (blue, cyan, orange, green)
curve is calculated at the disk midplane (1$H$, 2$H$, 3$H$, 4$H$).  
The shaded region represents the region where the density perturbation is determined by the planet and the circumplanetary region instead of the spiral shock. } \label{fig:denmaxmin}
\end{figure*}

\begin{figure*}[ht!]
\centering
\includegraphics[trim=0cm 1.8cm 0cm 1.2cm, width=1.0\textwidth]{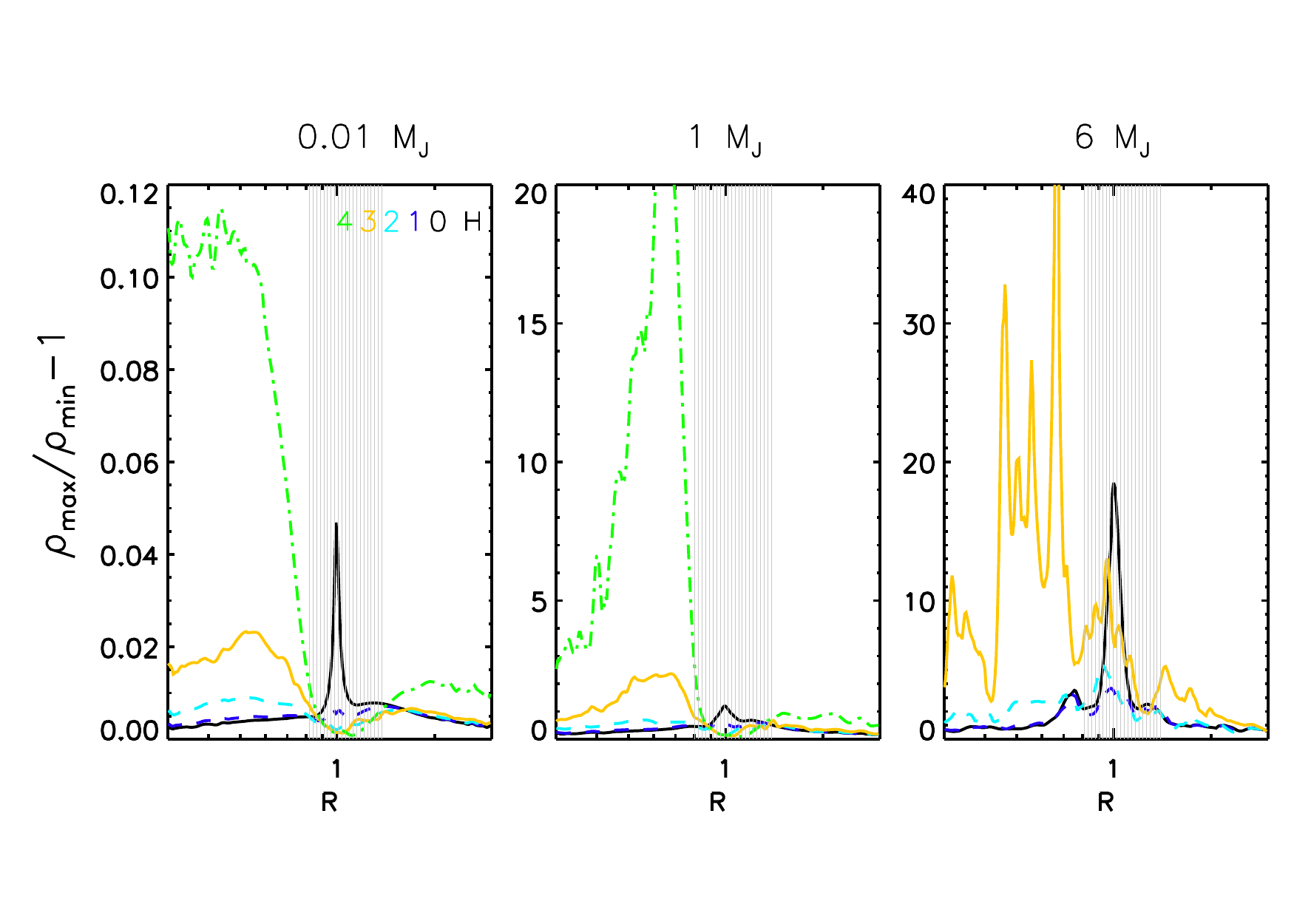} 
\vspace{0.0 cm}
\caption{Similar to Figure \ref{fig:denmaxmin} but for SM1T1, SM2T1, and SM3T1.} \label{fig:denmaxmintc1}
\end{figure*}

Finally, to illustrate the increase of the density perturbation with height in disks and the qualitative difference between
inner and outer arms, we plot the relative density perturbation along the radius
at different heights ($z=$0, 1, 2, 3, 4 $H$) in Figures \ref{fig:denmaxmin} and \ref{fig:denmaxmintc1}. 
The relative density perturbation is defined as
$\rho_{max}(R,z)/\rho_{min}(R,z) -1$, where $\rho_{max}(R,z)$ and $\rho_{min}(R,z)$ are the maximum and minimum density
along the azimuthal direction ($\phi=[0, 2\pi]$) at the fixed $R$ and $z$. We can see that the inner arms and outer arms are qualitatively different. 
For inner arms, the relative density
perturbation is getting larger at higher altitudes, and the  perturbation  can increase by more than a factor of 10 from the midplane
to the disk surface. For the outer arms, the relative density perturbation is almost unchanged between the midplane and the disk surface in isothermal disks
(Figure \ref{fig:denmaxmin}) and only increases slightly from the midplane to the disk surface 
in non-isothermal disks
(Figure \ref{fig:denmaxmintc1}). Overall, for the inner arms, the large density perturbation at the disk
surface has a significant effect on the near-IR observations as shown below.

\section{Discussion}

\subsection{{Numerics,  Longterm Evolution, and Different Disk Structures}}
 \begin{figure}[ht!]
\centering
\includegraphics[trim=0.8cm 1.0cm 0.2cm 0.2cm, width=0.5\textwidth]{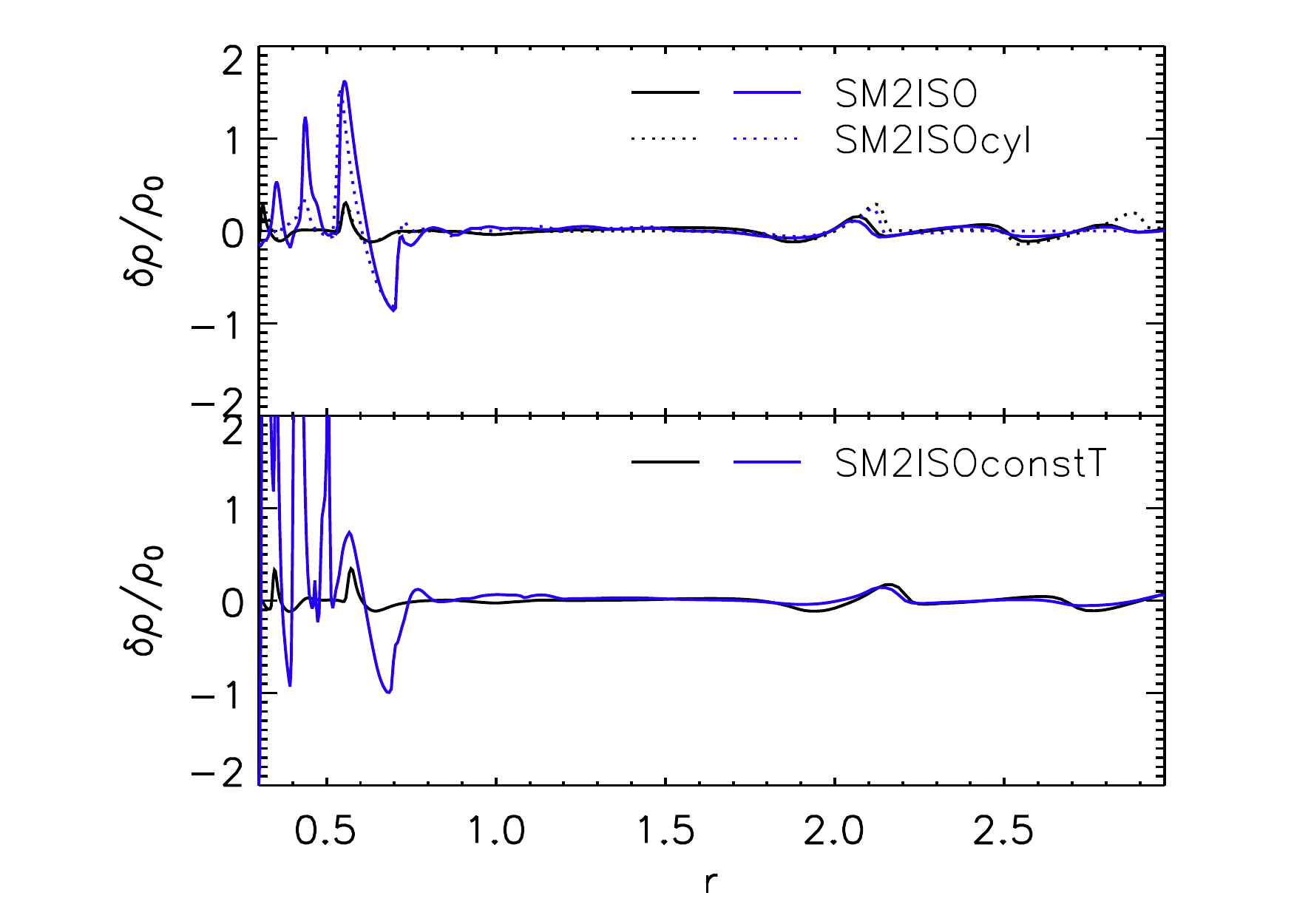} 
\vspace{0.0 cm}
\caption{ The relative density perturbation at the disk midplane (black curves) and at $\theta=\pi/2+0.35$ (cyan curves) for SM2ISO (solid curves in
the upper panel), SM2ISOcyl (dotted curves in the upper panel), and SM2ISOconstT (solid curves in the lower panel). } \label{fig:oned}
\end{figure}

As shown in Figure \ref{fig:shockvender}, the 3-D spiral shocks are not perpendicular to the disk midplane.  They curl in a way 
that they are almost along the $\theta$ direction in the spherical-polar grid.  To quantify the curl, the upper panel of Figure \ref{fig:oned} 
shows the relative density perturbation at the disk midplane (black curves) and at $\theta=\pi/2+0.35$ (cyan curves) for run SM2ISO (solid curves). 
Please note that the x-axis in Figure \ref{fig:oned} is $r$ (the radial position in spherical-polar coordinates) instead of $R$ (the radial position in cylindrical coordinates). 
The fact that the shock density peaks at almost the same $r$ at either the disk midplane or $\theta=\pi/2+0.35$ demonstrates that the shocks almost curl
along the $\theta$ direction. Such coincidence 
 makes us suspect that they could be numerical artifacts due to the adopted spherical-polar grid structure. 
 Thus, we use Athena to carry out a similar simulation but using  cylindrical coordinates (run SM2ISOcyl introduced in \S 3.3). 
 After transforming  simulation outputs from cylindrical coordinates to spherical-polar coordinates, 
the relative density perturbation for SM2ISOcyl is shown in the upper panel
 of Figure \ref{fig:oned} as the dotted curves.  The  shocks are at similar positions as those in SM2ISO, which 
 confirms that the curled shock structure is real instead of numerical artifacts.

The physical reason for the curled shock structure is unclear. One would suspect that it may be related to the radial temperature gradient in the local isothermal disk.
In such disks, the disk rotates at different angular velocities at different disk heights. Such vertical shear can change disk dynamics, such as leading to
the vertical shear instability (Nelson \etal 2013) and it may also lead to curled shock structures. 
To test this idea, we have run one simulation with a constant temperature in the whole disk (SM2ISOconstT). In such a disk,
the disk rotates at a constant angular velocity at a given $R$ independent on the disk height. The relative density perturbation is shown in the bottom panel of Figure \ref{fig:oned}.
However, the shock density still peaks at almost the same $r$ at different $\theta$, implying that the curl of the shock is not caused by the vertical shear in disks. 
Thus, the physical mechanism for the curled shock remains unclear and still deserves further explore.

 \begin{figure}[ht!]
\centering
\includegraphics[trim=0.8cm 1.0cm 0.2cm 0.2cm, width=0.5\textwidth]{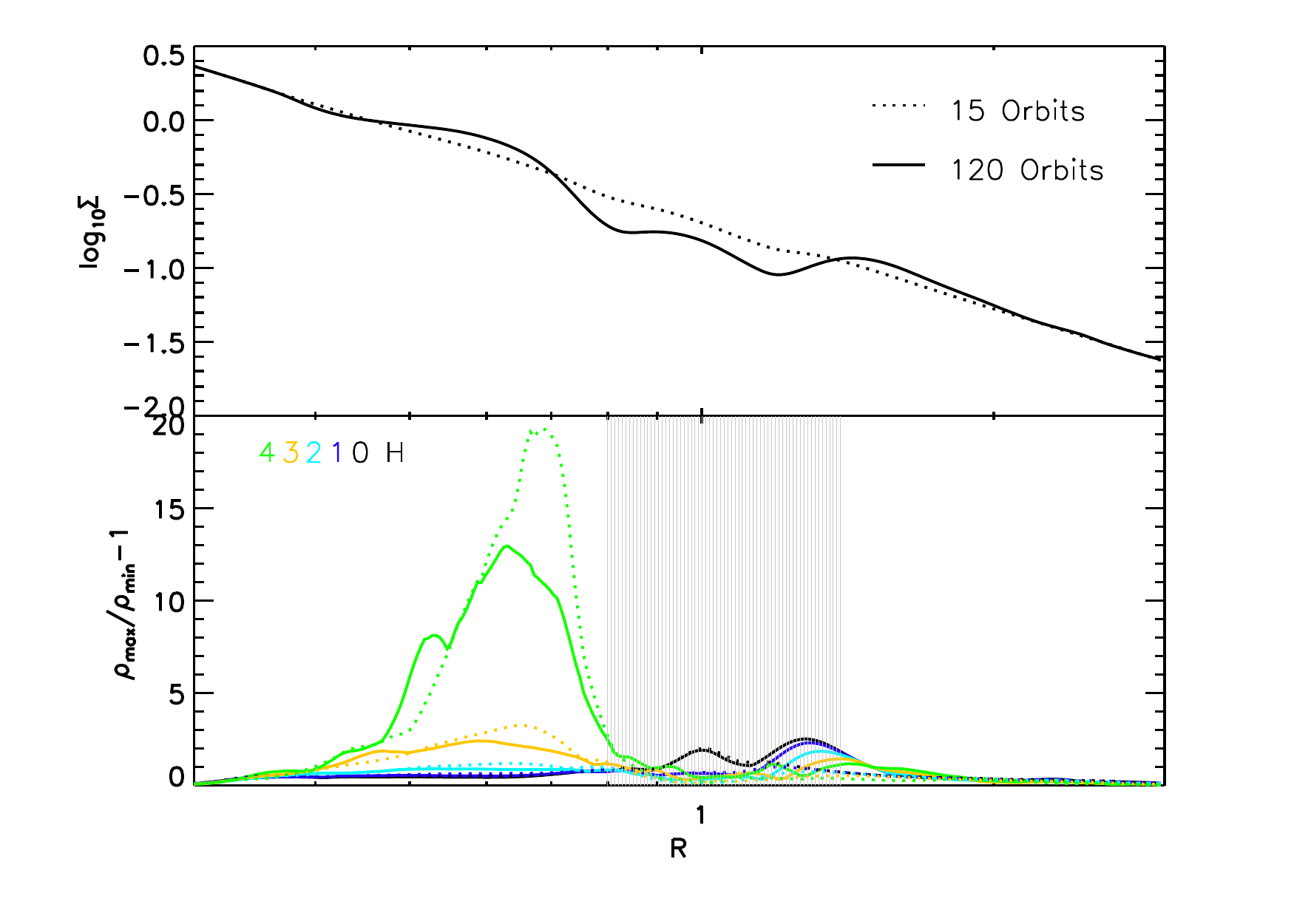} 
\vspace{0.0 cm}
\caption{The disk surface density (upper panels) and density perturbation (lower panel) at 15 (dotted curves) and 120 (solid curves) orbits for run SM2ISOlong. 
Different colors in the lower panel represent  density perturbation at different disk heights similar to Figure \ref{fig:denmaxmin}.  The disk surface density
perturbation ($\Sigma_{max}/\Sigma_{min}$-1) is very similar to  the density perturbation ($\rho_{max}/\rho_{min}$-1) at the disk midplane.} \label{fig:longterm}
\end{figure}

To demonstrate  that our derived shock structure is independent on specific numerical choices in 
our model, we show the density perturbation for run SM2ISOlong in Figure \ref{fig:longterm}.  Run SM2ISOlong is different from SM2ISO
in several aspects (as introduced in \S 3.3) : 1) the simulation runs for 120 orbits instead of 10 orbits; 2) 
it has a wave damping region, 3) the planet mass is ramped up slowly over 10 orbits, and 4) it uses zero viscosity.  Despite these differences, its
 density perturbation at 15 orbits (bottom panel of Figure \ref{fig:longterm}) is quite similar to the density perturbation in SM2ISO (middle panel of Figure \ref{fig:denmaxmin}).
When a gap is induced at 120 orbits, the spiral shocks become weaker at the disk region close to the gap, while it still maintains the same
strength at the region far away from the gap.  We notice that vortices start to develop at the outer gap edge ($R\sim 1.4$) at 120 orbits, which 
slightly enhances the density perturbation at the outer gap edge. 

 \begin{figure}[ht!]
\centering
\includegraphics[trim=0cm 0.0cm 0cm 0cm, width=0.5\textwidth]{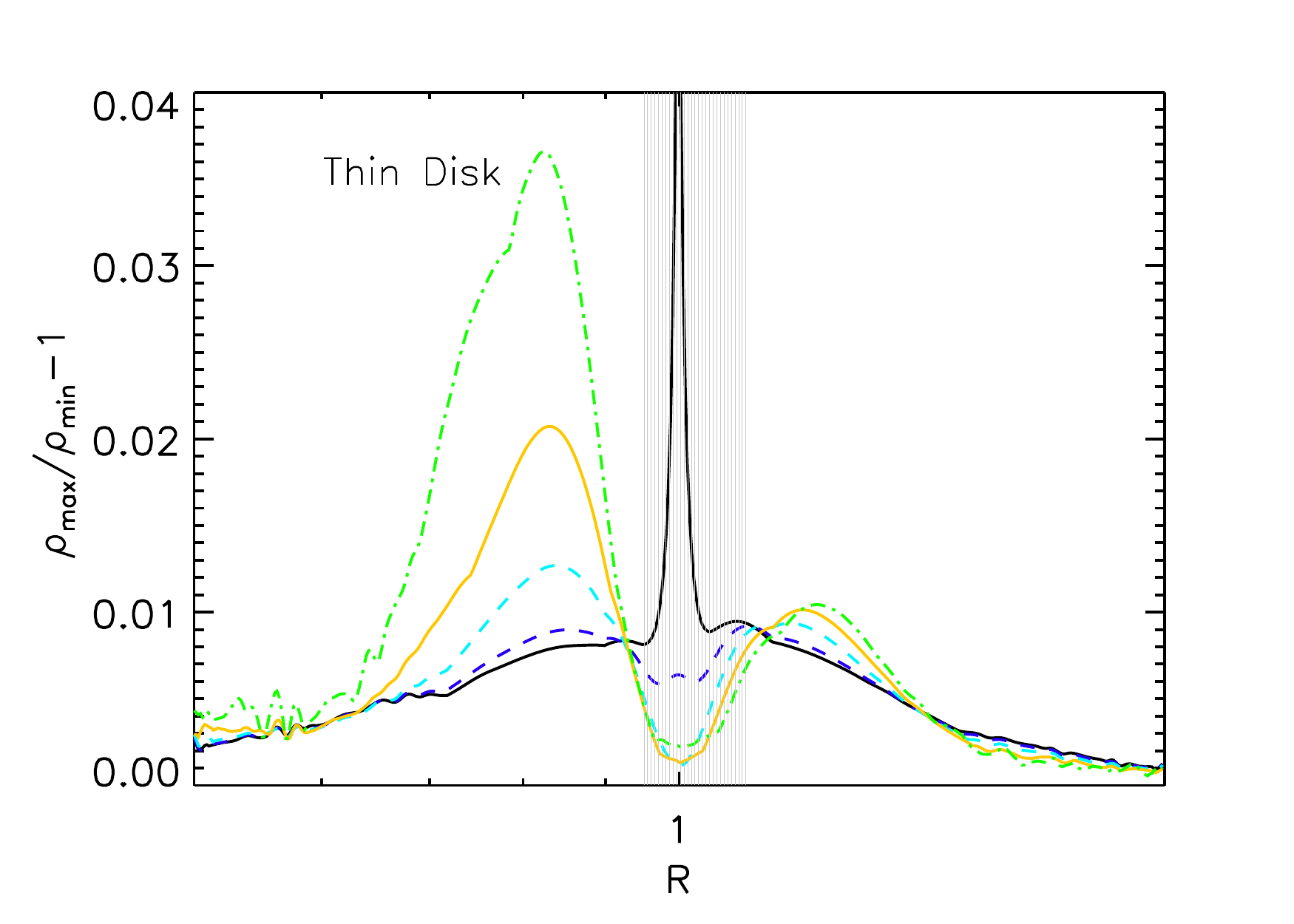} 
\vspace{0.0 cm}
\caption{ Density perturbation for run STHIN.  Different colors represent  density perturbation at different disk heights similar to
Figure \ref{fig:denmaxmin}. } \label{fig:thindisk}
\end{figure}

We have also studied how the shock structure can be affected by different disk structures.
Figure \ref{fig:thindisk} shows the relative density perturbation of the wakes in a thinner disk (STHIN).
Compared with Figure \ref{fig:denmaxmin}, we can see that the density perturbation increases by
a factor of 2 from the midplane to 3 $H$, and a factor of 4 from the midplane to 4 $H$ in the thin disk,
compared with a factor of 4 and 10 respectively in the thick disk. 
This suggests that the wakes/shocks have more significant 3-D structure in a thicker disk.
Although this does not favor detecting spiral arms in thinner disks in future, 
we need to keep in mind that the thermal mass (Equation  \ref{eq:thermal})  is smaller in a thinner disk so that
  the same mass planet corresponds to a more massive planet in the scale of thermal mass and it will excite stronger
density waves. In this case the inner arms may still be observable in a thin disk.

\subsection{Near-IR Images}
\begin{figure*}[ht!]
\centering
\includegraphics[trim=0cm 0.8cm 0cm 0cm, width=1.0\textwidth]{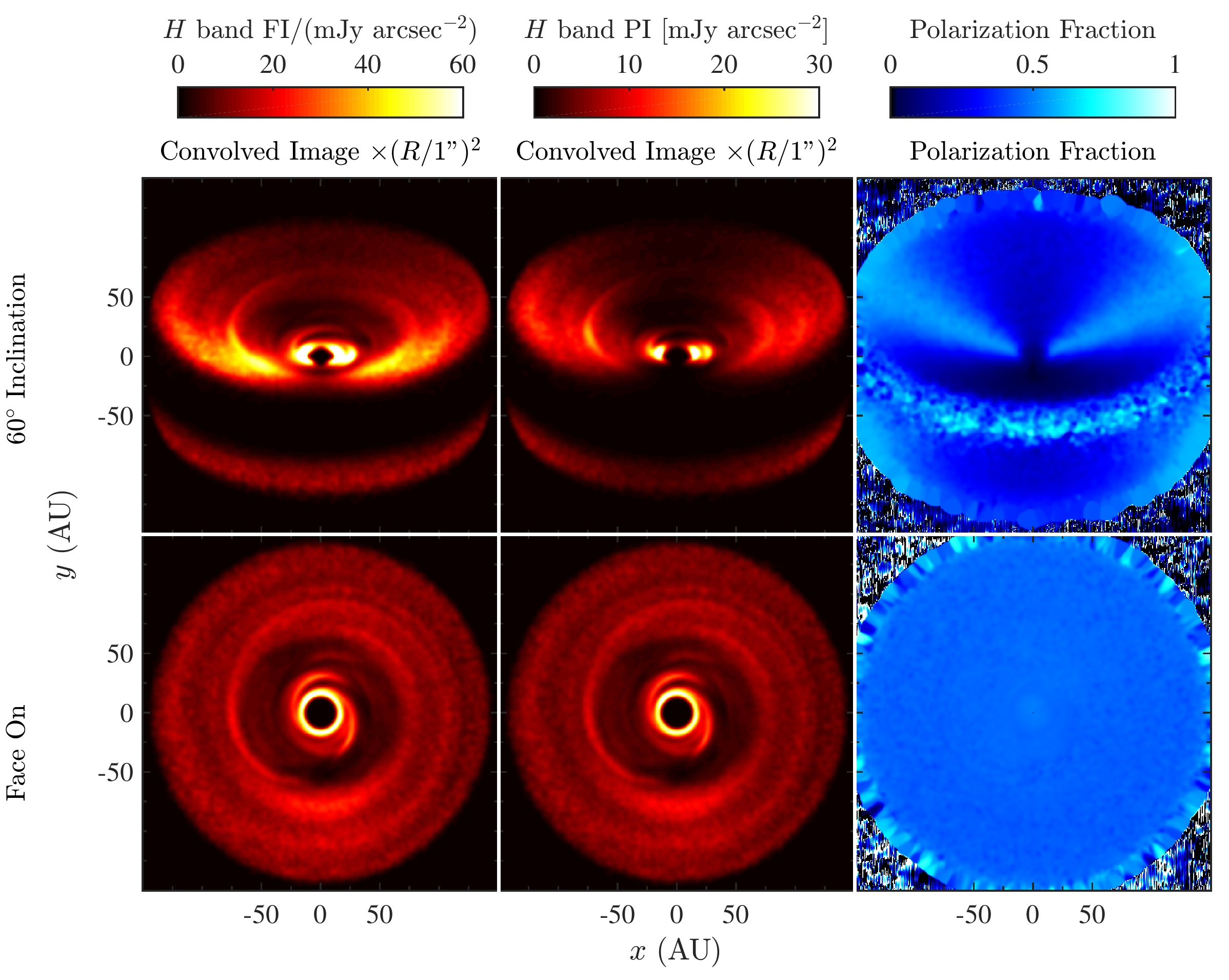} 
\vspace{0.0 cm}
\caption{Full intensity  (left panels) and polarized intensity {after scaled to R$^{-2}$} (middle panels), and polarization fraction (right panels) 
for SM3ISO ($M_{p}=6 M_{J}$) when the disk is viewed at the 60$^{o}$ inclination angle 
(upper panels) and face on (bottom panels). } \label{fig:image6mjfull}
\end{figure*}

\begin{figure*}[ht!]
\centering
\includegraphics[trim=0cm 0.8cm 0cm 0cm, width=1.0\textwidth]{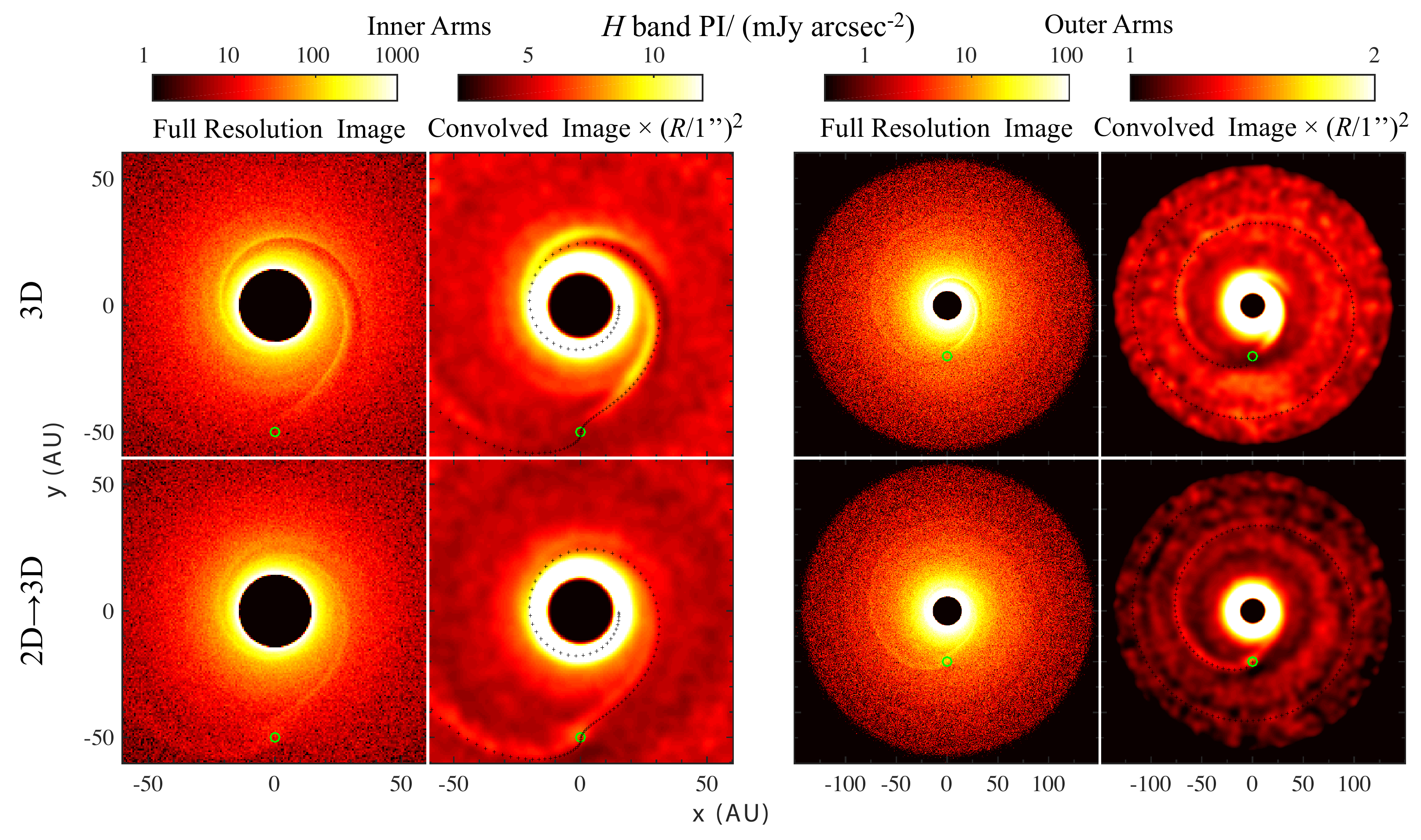} 
\vspace{0.0 cm}
\caption{The near-IR  polarized intensity maps for inner (left two panels) and outer (right two panels) arms for SM2ISO ($M_{p}=M_{J}$). 
The planet is assumed at 50 AU. In the left two panels, we assume the system is 70 pc away so that the inner arms are shown clearly, 
while in the right two panels we assume the system is 140 pc away.
The full resolution images have been convolved with a 0.06'' beam to derive the convolved images.
The upper panels show the images using disk structure directly from 3-D simulations while the bottom panels 
use disk structure assuming the disk is in vertical hydrostatic equilibrium. The dotted curves are the positions of the spiral wake
derived from linear theory (Equation \ref{eq:linearwake}). 
} \label{fig:image1mj}
\end{figure*}

\begin{figure*}[ht!]
\centering
\includegraphics[trim=0cm 0.8cm 0cm 0cm, width=1.0\textwidth]{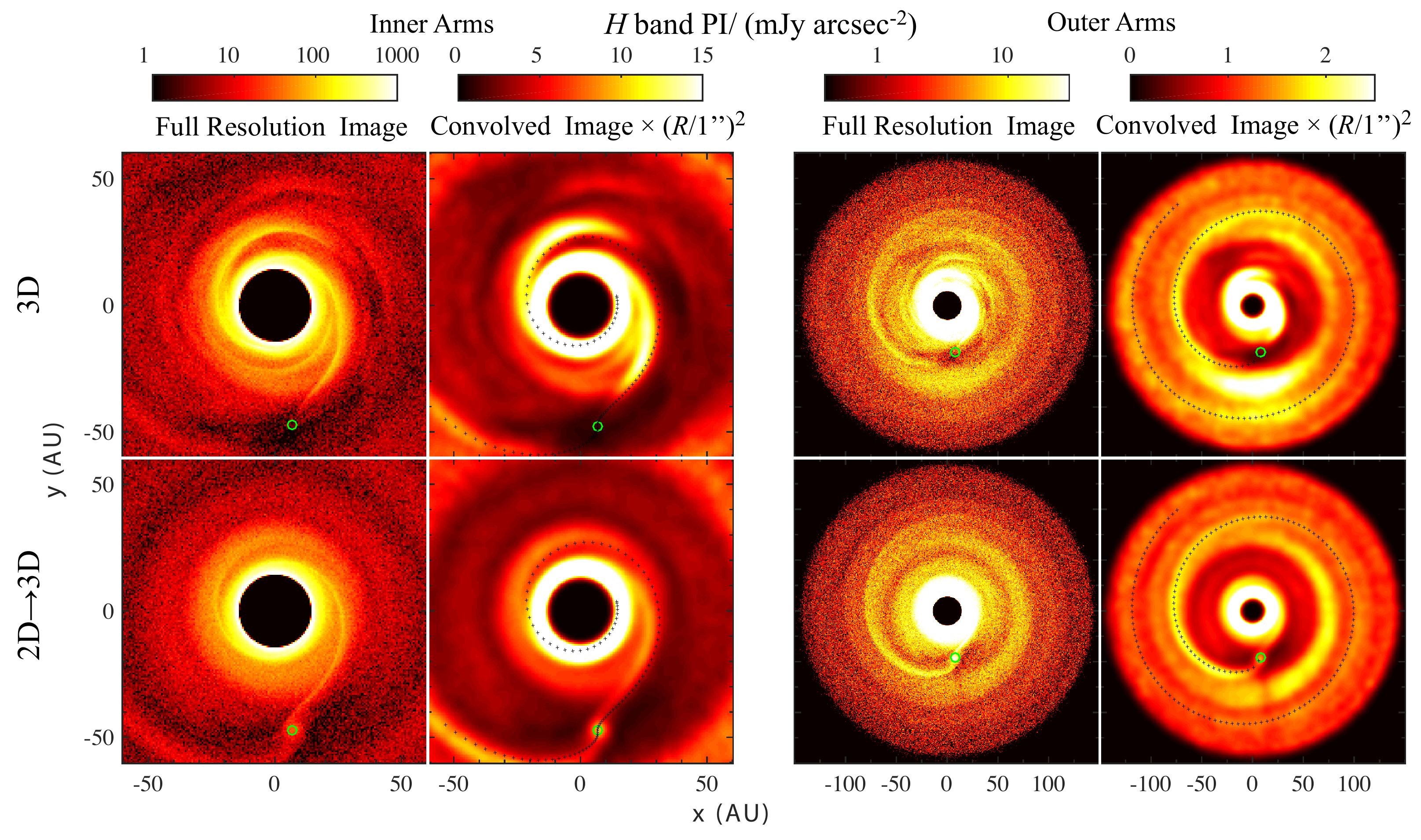} 
\vspace{0.0 cm}
\caption{Similar to Figure \ref{fig:image1mj} but for SM3ISO ($M_{p}=6 M_{J}$). } \label{fig:image6mj}
\end{figure*}

To understand how the 3-D structure of density waves/shocks affects observations, we post-process our hydrodynamical simulations with
 Monte-Carlo radiative transfer calculations to generate near-IR scattered light images (Figures \ref{fig:image6mjfull}, \ref{fig:image1mj} and
 \ref{fig:image6mj}). 
 The details on the Monte-Carlo radiative transfer calculations are presented in Dong \etal (2014, 2015). 
 To assign physical scales to our simulations, we assume that the planet is at 50 AU and 
 the central source is a typical Herbig Ae/Be star ({2 M$_{\odot}$}) with a temperature
of 10$^4$ K and a radius of 2$R_{\odot}$. ISM dust grains have been used and their distribution
 is assumed to follow the gas distribution. {Following Dong \etal (2015), we assume that the gas disk is 0.02 $M_{\odot}$. 
 With the gas to dust mass ratio of 100:1, the total dust mass is 2$\times10^{-4}M_{\odot}$.
 Further assuming 10\% of dust is in the form of ISM dust, the total mass
 of the ISM dust is 2$\times10^{-5}M_{\odot}$. In this model, Toomre Q parameter at 50 AU is $\sim$30,   so neglecting disk self-gravity is justified.}
 In MCRT simulations, photons from the central star are absorbed/reemitted or scattered by
the dust in the surrounding disk.  The convolved images in Figures \ref{fig:image6mjfull}, \ref{fig:image1mj} and  \ref{fig:image6mj}
are derived by convolving
full resolution  images with a Gaussian point spread function having a full width half maximum (FWHM) of 0.06''.
This resolution is comparable with NIR
direct imaging observations using Subaru, VLT, and Gemini. 
In the right two panels of  Figures \ref{fig:image1mj} and
 \ref{fig:image6mj}, we assume that the object is 140 pc away, while 
 in Figure \ref{fig:image6mjfull} and the left two panels of  Figures \ref{fig:image1mj} and
 \ref{fig:image6mj} we assume that the distance is 70 pc so that the inner arms are shown more clearly. 
 
{Both full intensity and polarized intensity images are calculated and shown in Figure 
\ref{fig:image6mjfull}. The spiral arms are evident in these images.
When the disk is viewed face-on, both full intensity and polarized intensity images
are almost identical except that the full intensity image is almost a factor of 2 brighter than the polarized intensity image.
The polarization fraction, which is defined as the ratio between the polarized intensity
and full intensity, is almost a constant ($\sim$0.45). When the disk is viewed at some inclination
angle, the polarization fraction is not a constant due to the dust forward scattering. The dependence of the polarization fraction on disk inclination
can be useful to determine the disk inclination angle.}

 To highlight the importance of the 
 3-D wave structure, we have also computed models by only using the disk midplane density from
 simulations,
which is labeled as 2D$\rightarrow$3D in Figures \ref{fig:image1mj} and
 \ref{fig:image6mj}. 
{In these models, we assume that the disk is in vertical hydrostatic equilibrium and puff up the midplane
 density to higher altitudes as $\rho=\rho_{mid}{\rm exp}(-z^2/2h^2)$ where $h/R=0.1R^{0.25}$ (the same as the scale height used in 
 3-D simulations).}

 \begin{figure}[ht!]
\centering
\includegraphics[trim=0cm 0.8cm 0cm 0cm, width=0.5\textwidth]{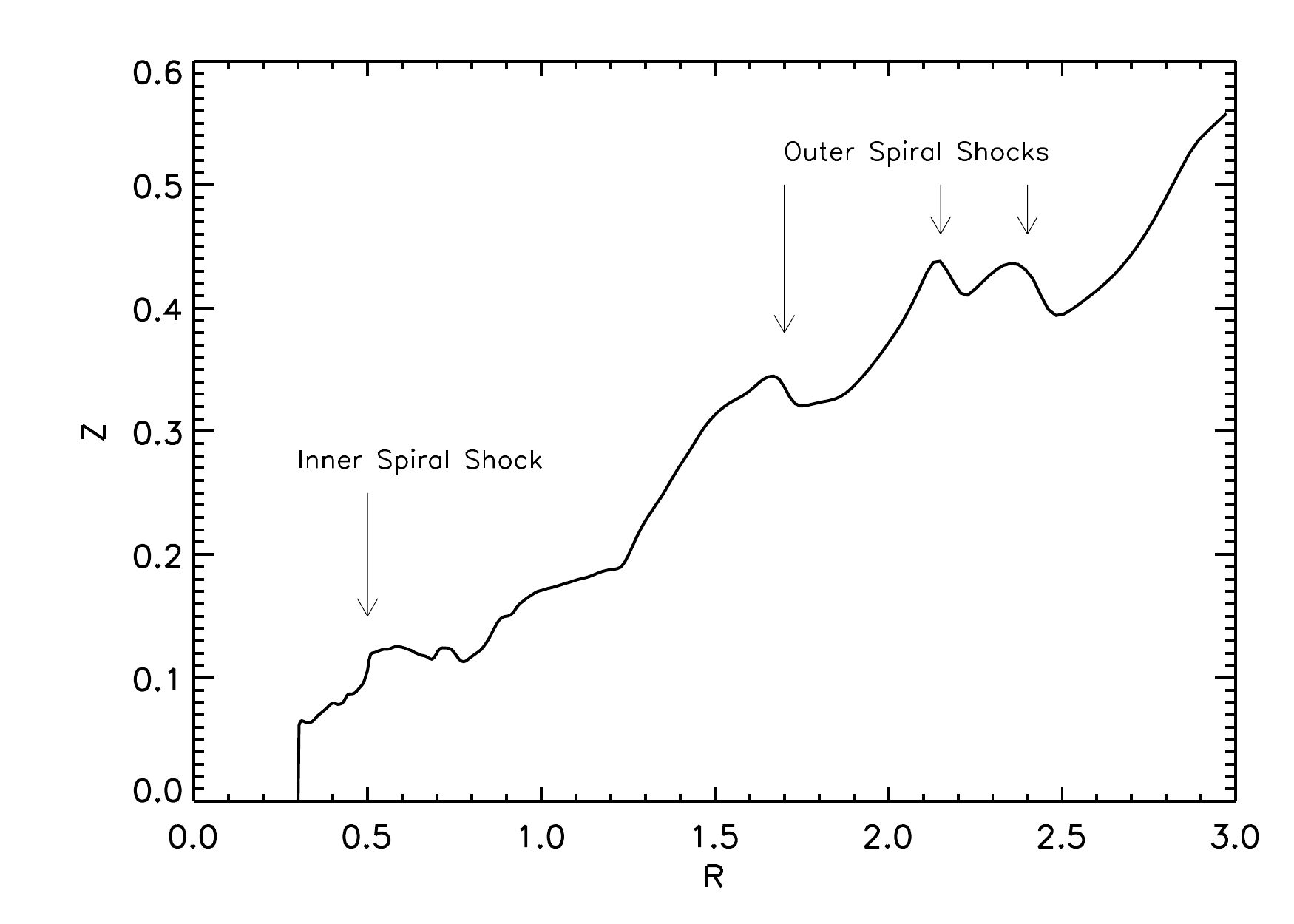} 
\vspace{0.0 cm}
\caption{ The scattering surface where the column density is 0.01 (code units) for SM3ISO.
Shock fronts are facing the star for the inner spiral shocks while the rarefaction waves are facing the star for the outer
spiral shocks. } \label{fig:scattering}
\end{figure}

Figures \ref{fig:image1mj} and
 \ref{fig:image6mj} show that, in 3D models, the inner arms are considerably 
 more prominent than the outer arms, and normally 
 a secondary arm can be as bright as the primary arm. The shape of the inner arms clearly deviate from the prediction of linear theory. As expected from the non-linear expansion of
spiral shocks, the pitch angle of the inner spiral arms in the more massive planet case (SM3ISO, Figure \ref{fig:image6mj})
is larger than those in  the less massive planet case (SM2ISO , Figure \ref{fig:image1mj}).
 The inner arms are also quite sharp, 
 while the outer arms are quite broad and sometimes indistinguishable from the background disk. 
 This difference is partly because the sharp shock fronts are facing the star for the inner 
 arms, while they are facing away from the  star for the outer arms.  {The different geometry at the disk surface can greatly affect
 the intensity of the scattered light images (Takami et al. 2014).}
 We calculate an approximate scattering surface, defined as the disk surface where the column density is 0.01 (in code  units), for the SM3ISO model at $\phi-\phi_{p}=78^{o}$,  shown in Figure \ref{fig:scattering}. Clearly, for the inner arms, the shock fronts are facing the star, while, for the outer arms, the 
smooth rarefaction waves  are facing the star. Since the rarefaction waves change
 gradually with radius, they are illuminated by the star more uniformly than the shock. Thus, the
 outer arms appear quite broad. However, when the planet mass is not very high (1 M$_{J}$ case),
 the width of rarefaction waves in the radial direction can be smaller than the size of the observational beam, and we won't be able to distinguish the inner and outer arms based on the sharpness of the arms.

 In 3-D models of Figure \ref{fig:image1mj} and \ref{fig:image6mj}, the secondary inner arms are as apparent as the primary arms, even though the
 primary arms have higher surface density than secondary arms.
 This is due to the corrugated motion discussed above in Figure \ref{fig:onedr0p5} and \ref{fig:vyr0p5}, which increases
 the density of the secondary arms at the disk surface. 
 The secondary arm is offset from the primary arm 
with some azimuthal angle, as also shown in the surface density plot (Figure \ref{fig:shock}). This offset is smaller in SM1ISO (Figure \ref{fig:image1mj})  than
that in SM6ISO (Figure \ref{fig:image6mj}). The two
spiral arms are $\sim100^{o}$ apart in SM1ISO (Figure \ref{fig:image1mj}), while almost 180$^{o}$ apart in the more massive planet case (SM6ISO, Figure \ref{fig:image6mj}).

By comparing 3D models with 2D$\rightarrow$3D models in Figure \ref{fig:image1mj} and \ref{fig:image6mj}, we find that the inner spiral arms are
more prominent in 3D models, as expected due to
 inner arms' higher density perturbation at the disk surface in 3D models (Figure \ref{fig:denmaxmin}). Even in convolved images,
 the polarized intensity of inner arms is at least twice stronger in 3D models than in 2D$\rightarrow$3D models.

The outer shocks  are not very apparent, and they are similar between 3D and 2D$\rightarrow$3D models since the density perturbation of outer arms is almost height independent (\S 5). As discussed in \S 4, the outer arms coincidently follow linear theory (Equation \ref{eq:linearwake}).
When the planet mass is very large as in Figure \ref{fig:image6mj}, the secondary outer arm starts to become visible .

 Another noticeable difference between 3D and 2D$\rightarrow$3D models is that the planet (or the circumplanetary region)
is bright in  2D$\rightarrow$3D models while it is  dim in 3D models. This is because, when we puff the  disk from 2-D to 3-D,
we have ignored the planet's gravity so that  the higher density in the circumplanetary region leads to the higher density at the disk surface.
The circumplanetary region even casts a shadow to the outer disk in 2D$\rightarrow$3D models. 
In realistic 3-D models, the gravity of the planet has been self-consistently included, which pulls the circumplanetary material towards the disk midplane and leads
to a lower density at the disk surface.  Thus, the circumplanetary region receives less irradiation by the central star, and becomes dark.  On the other hand, the outer disk
beyond the planet is better illuminated and thus becomes bright instead of being shadowed. 

However, we need to keep in mind that we have ignored the luminosity from the planet and the circumplanetary disk in our models. Zhu (2015) point out that accreting circumplanetary disks can be
very bright ($\sim$0.001 $L_{\odot}$ if the circumplanetary disk accepts onto Jupiter at
a rate $\sim 10^{-8}M_{\odot}/yr$).
Such high luminosity may be able to illuminate the circumplanetary region significantly. We may
also be able to directly detect such accreting circumplanetary disks in direct imaging observations 
operating at mid-IR wavelengths.

\subsection{{Observational Implications}}
Previous works suggest that spiral patterns from direct imaging observations cannot be explained by planet-induced spiral arms, since the  
 pitch angle of the spiral arm is larger in observations than that 
 predicted by the linear spiral wave theory, and the contrast of the spiral arm is higher in observations than suggested by the
 synthetic observations based on two dimensional planet-disk simulations. 
However, most previous works only focus on using outer spiral arms beyond the planet to explain observed spiral patterns. As shown in Figure \ref{fig:pitch},
the inner arms generally have larger pitch angle than outer arms even under the linear density wave theory. Furthermore, when the spiral
arms become spiral shocks, the pitch angle increases with stronger shocks.
We also found that a secondary (or even a tertiary) spiral arm, especially for the inner arms, is also excited by a massive planet. 
The more massive is the planet, the 
larger is the separation in the azimuthal direction between the primary and secondary arms. 
The inner arms also have significant  vertical motion, which boosts the density perturbation at the disk surface and the intensity contrast in synthetic images.

Thus, we have three independent ways to estimate the planet mass based on the spiral patterns in observations: 1)
using the deviation between the measured pitch angle and the pitch angle predicted in the linear theory (Figure \ref{fig:pitch});
2) the existence of the secondary spiral arm and the separation between two arms (Figure \ref{fig:shock}); 3) the intensity contrast of the spiral arms in observation (Figure \ref{fig:image1mj} \& \ref{fig:image6mj}).
The first two methods only use the information on the shape of the spiral arms and 
are less likely to be affected by the disk vertical temperature structure.
In a companion paper (Dong \etal 2015), we have shown that all these three methods suggest that
SAO 206462 and MWC 758 may harbor massive planets (several to tens of Jupiter masses) outside the detected spiral arms.

The planet-induced spiral arms can efficiently scatter light from the central star and
 are quite bright within gaps (Figure \ref{fig:image6mjfull}). The spiral arms may be responsible for
 some resolved infrared features within the cavity of LkCa 15 (Kraus 
\& Ireland 2012).

\section{Conclusion}
We have carried out two dimensional (2-D) and three dimensional (3-D) 
hydrodynamical simulations to study spiral wakes/shocks
excited by young planets.  Simulations with different planet masses (0.01, 1, and 6 $M_{J}$) and different equations of state (isothermal and adiabatic) have been carried out.

\begin{itemize}
\item  We find that the linear density wave theory can only 
explain the shape of the spiral wakes excited by a very low mass planet (e.g. 0.01 $M_{J}$).
Spiral shocks excited by high mass planets clearly deviate from the prediction of linear theory. 
For a more massive planet, the deviation is more significant and the pitch angle 
of the spiral arms becomes larger.
This phenomenon can be nicely explained by the wake broadening 
from the non-linear density wave theory (Goodman \& Rafikov 2001, Rafikov 2002). A more massive planet excites
a stronger shock which expands more quickly, leading to a larger pitch angle.

\item A secondary (or even tertiary) inner spiral arm is also excited by the planet. 
It seems to be excited at the edge of the N-shaped primary arm.
The more massive is the planet, the 
larger is the separation between the primary and secondary arm. 
At the disk surface,
the secondary inner arm can be as strong as the primary arm. 
The secondary inner arm almost keeps the same azimuthal separation with the primary arm
at every radius in the disk. {The excitation mechanism for the secondary (tertiary) arm deserves
further study.}

\item The spiral shocks have significant 3-D structure. They are not perpendicular
to the disk midplane. They are curled towards the star at the disk surface. This further
increases the pitch angle of the inner arms at the disk surface, but 
reduces the pitch angle of the outer arms at the disk surface. For outer arms, this effect compensates 
the increased pitch angle due to
wake broadening.  Eventually at the disk surface, 
the shape of outer spiral arms still roughly follows the prediction of linear theory,
while the inner arms are considerably more opened than predicted by the linear theory.

\item The inner spiral shocks also have significant vertical motion. The corrugated motion 
increases the density perturbation of the inner spiral arms by more than a factor of 10 at $z\sim 3-4 H$  compared with the perturbation at the disk midplane. This can dramatically increase the
contrast of the spiral patterns in near-IR scattered light images. The outer spiral shocks have little
vertical motion in isothermal disks. With a non-isothermal EoS, there are some vertical motions for the outer arms, which can make the outer arms more apparent. 

\item We have combined our hydrodynamical simulations with
 Monte-Carlo radiative transfer calculations to generate near-IR scattered light images.
 We find that the inner spiral arms are prominent features that are observable by current near-IR imaging facilities. 
Besides the apparent spiral patterns in some transitional disks, planet-induced spiral arms 
may also be responsible to some marginally detected near-IR features within cavity of some transitional disks. 
We further demonstrate that
inner spiral arms in synthetic near-IR images using full 3-D hydrodynamical models are much more prominent than those based on 2-D models assuming hydrostatic equilibrium, consistent with
the 3-D  structure of the inner arms. On the other hand, the outer shocks  are not very apparent and they are similar in synthetic images using 3D and 2D models. This indicates 
the need to model observations (especially for inner arms) with full 3-D hydrodynamics.

\item The different geometry between the inner and outer arms also affects their 
appearance in near-IR images. The sharp shock fronts of the inner arms face the
central star directly, producing sharp narrow spiral features in observations. On the other hand,
for the outer arms, the smooth rarefaction waves face the central star, producing broad and dimmer
spiral features.

\item In near-IR images, the circumplanetary region is very dim since the planetary gravity reduces
the density at the disk atmosphere. However, the disk region behind the planet can be better illuminated and becomes bright.

\end{itemize}

{In the Appendix, we have shown that buoyancy resonances are confirmed in global adiabatic 
simulations  even if the disk has a moderate cooling rate.
They can lead to sharp density ridges around the planet, which may have observational signatures.}

Overall, spiral arms (especially
inner arms) excited by low mass companions are prominent features
in near-IR scattered light images. Most importantly, 
we have three independent ways to infer the companion's mass: 1) the pitch angle of the spiral patterns, 2) the separation between the primary and secondary
arms, and 3) the contrast of the spiral patterns.

In a companion paper (Dong \etal 2015), we have combined MCRT and hydrodynamical 
simulations from this paper, and  shown that planet-induced inner arms
can explain spiral patterns revealed by recent near-IR direct imaging observations for SAO 206462 and MWC 758.
{We want to caution that our proposed model can not explain the gaps at the inner disks. Other mechanisms (e.g. another planet or
disk photoevaporation) have to be invoked to 
explain the inner gaps.}

\acknowledgments
All hydrodynamical simulations are carried out using computer supported by the 
Princeton Institute of Computational Science and Engineering, and the Texas Advanced Computing Center (TACC) 
at The University of Texas at Austin through XSEDE grant TG-
AST130002.
This
project is supported by NASA through Hubble Fellowship grants HST-HF-51333.01-A (Z.Z.)
and HST-HF-51320.01-A (R.D.) awarded by the Space Telescope Science Institute, which
is operated by the Association of Universities for Research in Astronomy, Inc., for NASA,
under contract NAS 5-26555. 

\appendix
\section{Buoyancy Resonances}
\begin{figure*}[ht!]
\centering
\includegraphics[trim=0cm 0.8cm 0cm 0cm, width=1.0\textwidth]{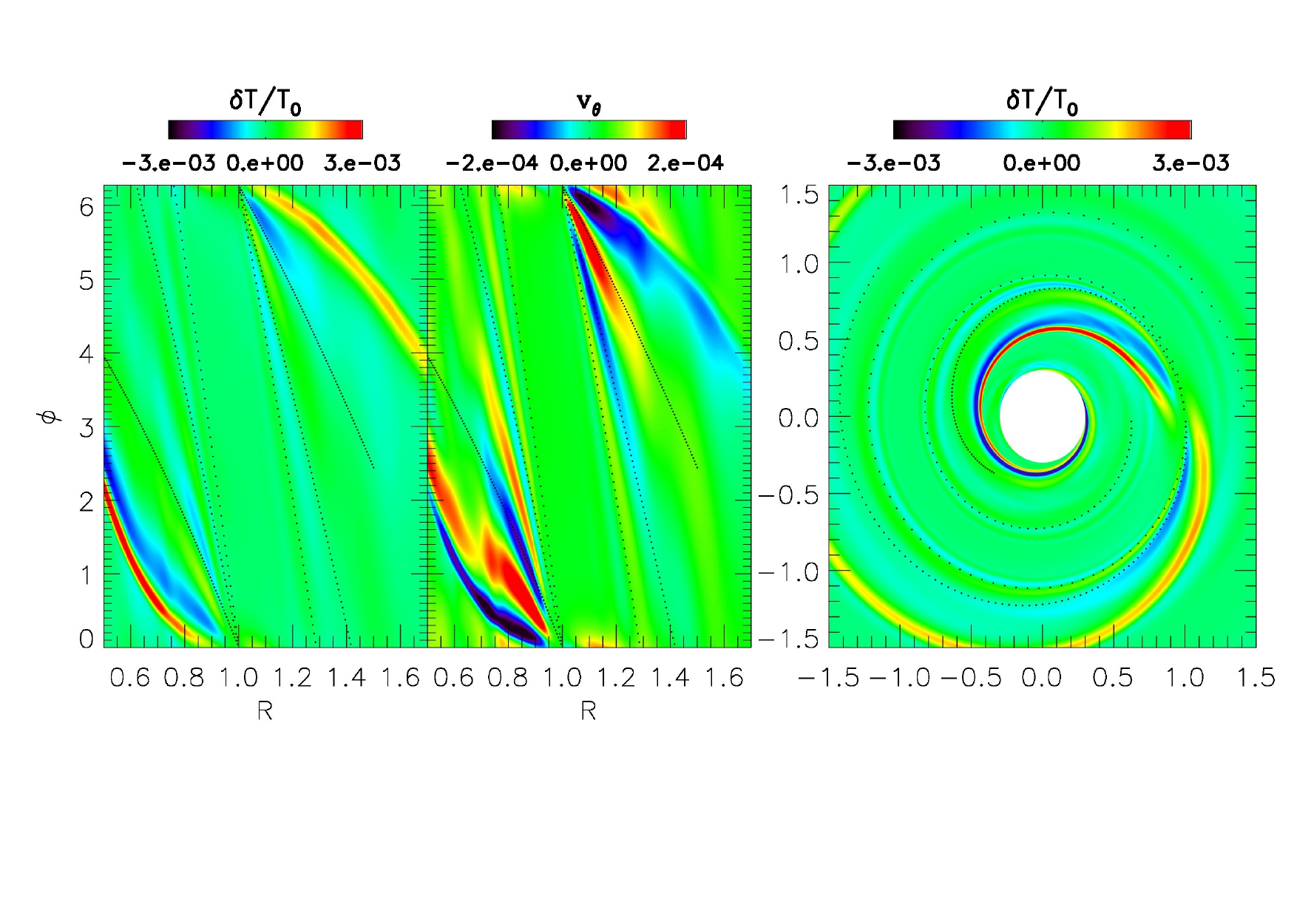} 
\vspace{0.0 cm}
\caption{Temperature fluctuations (the left panel) and $v_{\theta}$ (the middle panel) 
at $z=2 H$ for SM1T1. The right panel is the same as the left panel but in Cartesian coordinates. 
The dotted lines/curves are the position of buoyancy resonances from Equation (\ref{eq:buoyt}).  } \label{fig:buoy}
\end{figure*}

\begin{figure}[ht!]
\centering
\includegraphics[trim=0cm 0.8cm 0cm 0cm, width=0.5\textwidth]{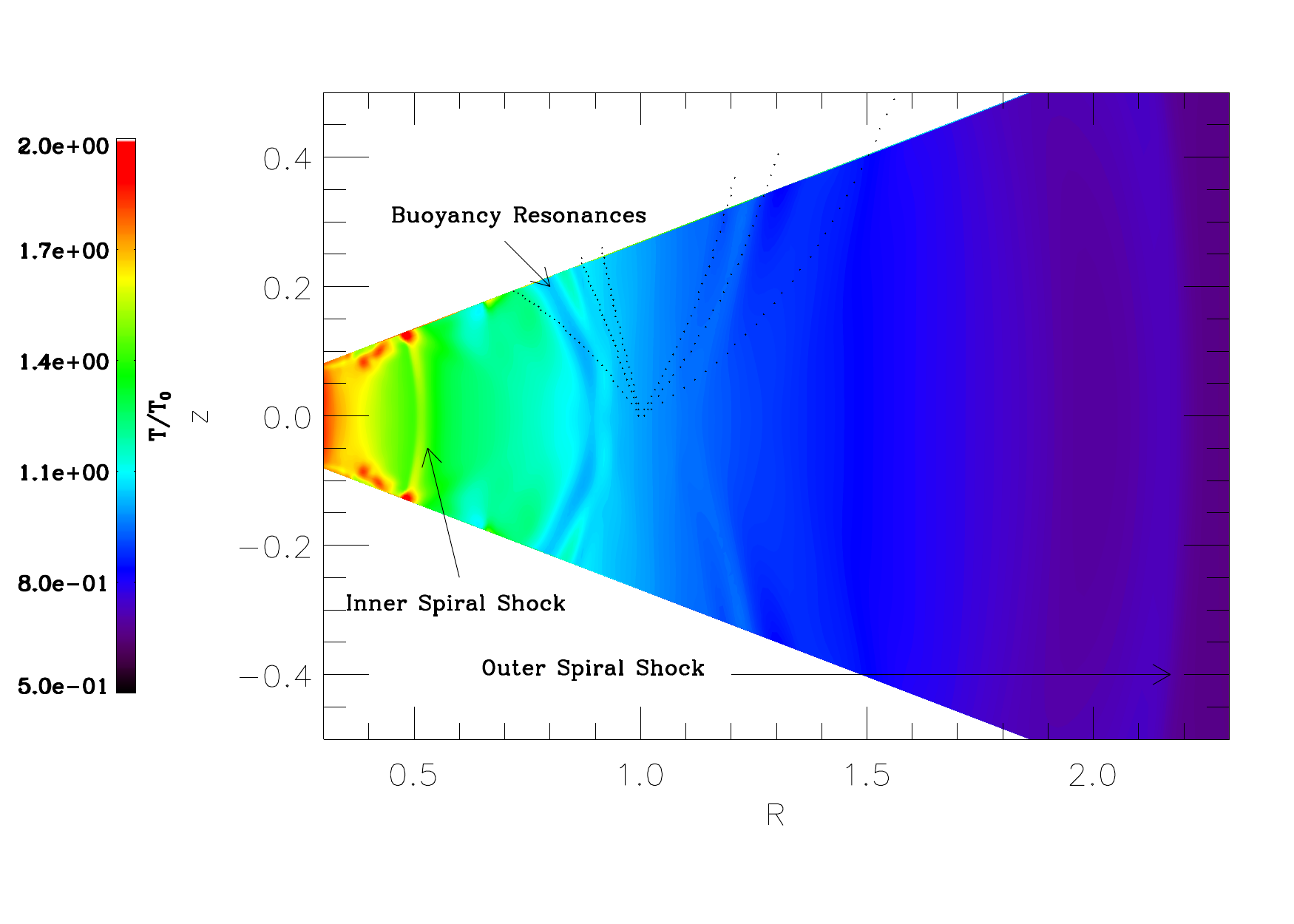} 
\vspace{0.0 cm}
\caption{The temperature structure for SM2T1 at an azimuthal slice ($\phi-\phi_{p}=100^{o}$).  Spiral
shocks and buoyancy resonances are labeled. The dotted lines are again from Equation (\ref{eq:buoyt}). } \label{fig:temp2D}
\end{figure}

The inner and outer spiral shocks are due to the steepening of spiral density waves which are excited by 
the planet at Lindblad resonances. At Lindblad resonances, the Doppler shifted frequency matches the disk
epicyclic frequency ($m(\Omega_{p}-\Omega)=\pm \kappa$) and density waves are excited. 

However, besides the epicyclic frequency, the disk also has other natural frequencies. When the disk is not strictly
isothermal, it has a non-zero Brunt-V\"ais\"al\"a frequency 
\begin{equation}
N(z)=\sqrt{\frac{\gamma-1}{\gamma}}\frac{g(z)}{c_{s,iso}}
\end{equation}
where $c_{s,iso}^2=p/\rho$.
Matching the Brunt-V\"ais\"al\"a frequency  with the Doppler shifted frequency, we have
\begin{equation}
\sqrt{\frac{\gamma-1}{\gamma}}\frac{\Omega_{K}(R) z }{H}\left(1+\frac{z^2}{R^2}\right)^{-3/2}=\pm m(\Omega_{p}-\Omega)\,.\label{eq:buoy}
\end{equation}
Given a $m$, Equation (\ref{eq:buoy}) gives the position of the resonances. 
These buoyancy resonances were discovered in shearing box simulations (Zhu, Stone, \& Rafikov 2012) 
and studied analytically in Lubow \& Zhu (2014). They have significant contributions to the 
planetary torque, especially around the planet, which may affect planet migration.  
These resonances are infinitely thin, and no waves are excited to carry the deposited angular momentum and energy away. Their dissipation relies on microscopic viscosity or radiative cooling.
Thin density ridges with large temperature and velocity variations appear at these resonances.  

When various $m$ modes overlap with each other, we can roughly estimate the position of the final density ridges caused by buoyancy resonances 
following Equation 10 and 11 in Zhu \etal (2012).  First, given a $m$, we can calculate the corresponding resonance position
at $R$ and $z$. Then using the azimuthal wavelength for this mode $\lambda=2\pi/m$,  the
geometric location of the constant phase $2n\pi$ ($n$ is integer) is given by $\phi=n\lambda$ (assuming  
the phase of buoyancy waves is 0 at the planet position.). Thus,
\begin{equation}
\phi=\pm 2n\pi (\Omega_{p}-\Omega)\sqrt{\frac{\gamma}{\gamma-1}}\frac{H}{\Omega_{K}(R)z}\left(1+\frac{z^2}{R^2}\right)^{3/2}\,. \label{eq:buoyt}
\end{equation}

We plot temperature fluctuations and $v_{\theta}$ at $z=2 H$ for SM1T1 in Figure \ref{fig:buoy}. The positions of 
buoyancy resonances given
by Equation (\ref{eq:buoyt}) are plotted with dotted lines/curves. Figure  \ref{fig:buoy}  shows that both temperature 
fluctuations and $v_{\theta}$ are nicely tracked by Equation (\ref{eq:buoyt}), suggesting that
buoyancy resonances exist in disks even with $T_{cool}=1$. 
  
Even though buoyancy resonances can affect planet migration, they may not be observed through direct imaging technique since
the density fluctuations caused by these resonances are much weaker than the spiral density waves excited by Lindblad resonances. 
 For example, in the $z=3H$ panels of Figure \ref{fig:shocktc1}, we can see some density fluctuations close to the corotation region (especially for SM2T1, the 1 $M_{J}$ case),
 but they are much weaker than the spiral shocks. 

These buoyancy resonances also have vertical structure, as shown in Figure \ref{fig:temp2D}.
We sliced through the disk at a fixed $\phi$, and
the buoyancy resonance curves are plotted as dotted curves in Figure \ref{fig:temp2D}.
At the disk midplane, $N=0$ and there are no buoyancy resonances. Since $N$ 
increases with disk height, $|\Omega_{p}-\Omega|$ also needs to increase with height to match
the Brunt-V\"ais\"al\"a frequency considering $\lambda$ and m are the same at the same $\phi$. Thus, the curves move away from the planet position towards
the disk atmosphere. 

Figure \ref{fig:temp2D} also shows the inner and outer spiral shocks which are hotter
at the shock position.

\end{document}